\documentclass[11pt]{article}

\usepackage{fancyhdr}

\usepackage{geometry}

\RequirePackage[numbers]{natbib}

\newcommand{\bc}{\begin{center}}
	\newcommand{\ec}{\end{center}}

\newcommand{\bit}{\begin{itemize}}
	\newcommand{\eit}{\end{itemize}}

\newcommand{\be}{\begin{eqnarray*}}
	\newcommand{\ee}{\end{eqnarray*}}
\newcommand{\ben}{\begin{eqnarray}}
\newcommand{\een}{\end{eqnarray}}

\newcommand{\g}{\,\vert\,}

\newcommand{\D}{\mathcal{D}}
\newcommand{\G}{\mathcal{G}}

\newcommand{\T}{\mathcal{T}}
\newcommand{\N}{\mathcal{N}}

\newcommand{\Y}{\mathcal{Y}}

\newcommand{\Gtau}{\mathcal{G}_{\tau}}

\newcommand{\pa}{\mathrm{pa}}
\newcommand{\fa}{\mathrm{fa}}

\newcommand{\neigh}{\mathrm{ne}}

\newcommand{\bA}{\bm{A}}

\newcommand{\bD}{\bm{D}}

\newcommand{\bN}{\bm{N}}

\newcommand{\btheta}{\bm{\theta}}

\usepackage{tikz}
\usetikzlibrary{graphs}
\usetikzlibrary{positioning}

\newcommand{\black}{\color{black}}

\usepackage{bm}
\usepackage{bbm}
\usepackage{float}
\usepackage{amssymb}
\usepackage{subfig}
\usepackage{enumerate}
\usepackage{longtable}
\usepackage{mathtools,cancel}
\usepackage{url}
\usepackage{multirow}
\usepackage{lscape}
\usepackage{authblk}

\begin{document}

\title{Equivalence class selection of categorical graphical models}
\author[1]{Federico Castelletti \thanks{federico.castelletti@unicatt.it}}
\author[2]{Stefano Peluso \thanks{stefano.peluso@unimib.it}}
\affil[1]{Department of Statistical Sciences, Universit\`{a} Cattolica del Sacro Cuore, Milan}
\affil[2]{Department of Statistics and Quantitative Methods, Universit\`{a} degli Studi di Milano-Bicocca, Milan}


\date{}




\maketitle

\begin{abstract}
Learning the structure of dependence relations between variables is a pervasive issue in the statistical literature.
A directed acyclic graph (DAG) can represent a set of conditional independences, but different DAGs may encode the same set of relations and are indistinguishable using observational data. Equivalent DAGs can be collected into classes, each represented by a partially directed graph known as essential graph (EG).
Structure learning directly conducted on the EG space, rather than on the allied space of DAGs, leads to theoretical and computational benefits. Still, the majority of efforts in the literature has been dedicated to Gaussian data, with less attention to methods designed for multivariate categorical data.
We then propose a Bayesian methodology for structure learning of categorical EGs. Combining a constructive parameter prior elicitation with a graph-driven likelihood decomposition, we derive a closed-form expression for the marginal likelihood of a categorical EG model. Asymptotic properties are studied, and an MCMC sampler scheme developed for approximate posterior inference. We evaluate our methodology on both simulated scenarios and real data, with appreciable performance in comparison with state-of-the-art methods.

\vspace{0.7cm}

\end{abstract}

\section{Introduction}
The wide spread of complex data has increasingly raised the interest of statisticians in the development of appropriate tools to investigate structured dependence relations between variables. In this context, graphical models represent a powerful methodology 
(\citet{Laur:1996}), with directed acyclic graphs (DAGs) particularly suitable for many scientific problems, expecially biological (\citet{Frie:2004,Sachs:Etal:2005,Shoj:Mich:2009,Nagarajan:2013}).
A DAG encodes a set of conditional independencies between variables which can be read off from the graph using various criteria such as \textit{d-separation} (\citet{Pear:2000}). 
While in some fields (e.g. genomics) such dependence relationships can be postulated \emph{a priori} based on experts' knowledge, realistically the underlying DAG is unknown and accordingly needs to be inferred from the data.
In the Bayesian framework this corresponds to a model selection
problem which requires first the specification of a prior distribution on the space of DAG models and parameters.
The latter, combined with the data likelihood, leads to an integrated (marginal) likelihood and in turn to a posterior distribution on the DAG space. In this direction, the methodology deployed by \citet{Geig:Heck:2002} for parameter prior construction implies desirable properties of the DAG marginal likelihood. 

An additional complication arises because different DAGs may encode the same set of
conditional independencies (Markov equivalent DAGs). Markov equivalence
therefore induces a partition of the DAG space into Markov equivalence classes (\citet{Ande:etal:1997}). Under common distributional assumptions all DAGs in the same Markov equivalence class are indistinguishable using obervational data (\citet{Pear:2000}), and therefore should have the same marginal likelihood, a requirement
known as \textit{score equivalence}. 
In addition, model selection algorithms that ignore Markov equivalence
can fall into incoherences and computational inefficiencies as pointed out by \citet{Ande:etal:1997}. 
All DAGs in the same Markov equivalence class can be represented by an essential graph
(EG, \citet{Ande:etal:1997}, \citet{Chic:2002}), a chain graph (CG) whose chain components are decomposable undirected graphs (UG) linked by arrowheads. 

Clearly, structural learning of EGs always guarantees score equivalence, since it operates at level of equivalence classes. In this context, first efforts
to the investigation of the EG space have been confined to small graphs (\citet{Gill:Perl:2002}),
whilst more recently larger graphs
were studied by \citet{Sonn:etal:2015} and \citet{He:etal:2013} using Markov chain Monte Carlo (MCMC) methods.
The recent work of \citet{Caste:etal:2018} relies on the method of \citet{Geig:Heck:2002} to construct a parameter prior for Gaussian EGs, following the objective Bayes perspective of \citet{Cons:Laro:2012} and \citet{cons:et:al:2017} for, respectively, Gaussian and covariate-adjusted DAG models.
On the frequentist side, \citet{Chic:2002} 
provides an EG estimate using a greedy equivalence search (GES) algorithm based on additions and deletions of single edges, later modified for better estimation by \citet{Haus:Buhl:2012}.
Moreover, \citet{Spir:Glym:Sche:2000} proposed the PC algorithm, a constraint-based method which implements a sequence of conditional independence tests. 

All the techniques above mentioned have been mainly designed for data whose nature justifies the Gaussian assumption.
Even if graphical models for categorical data (also called \emph{Bayesian networks}) are widely employed in many domains
(\citet{scutari2014bayesian}), to our knowledge the Bayesian literature on categorical EG learning is narrow, limited to \citet{Madi:Etal:1996} and \citet{Cast:Perl:2004}.
The adoption of Bayesian scores by frequentist score-based methods can partially fill this gap, but clearly does not represent a fully satisfying solution. 
Still on the frequentist side, the PC algorithm of \citet{Spir:Glym:Sche:2000} can be adapted for categorical data and provides an EG estimate through a sequence of conditional independence tests, also adopted for contingency tables in \citet{scutari2014bayesian}. \citet[Chapter 9]{korb2010bayesian} and \citet[Chapter 16]{murphy2012machine} extensively cover categorical structure learning, but none of them at the level of equivalence classes. Also, the score-based algorithms outlined in \citet{scutari2018} only learn discrete DAGs. Furthermore, with categorical data different hyperprior specifications lead to different common scores, with crucial impact on the performance, and may easily compromise score equivalence. 

In the present paper we propose a fully Bayesian structure learning method for categorical EGs. Following the approach of \citet{Geig:Heck:2002}, originally introduced for DAG models, we derive a closed-form expression for the marginal likelihood of an EG, therefore avoiding any issue related to lack of score equivalence or hyperprior misspecification. 
Exploiting the Markov chain in \citet{He:etal:2013} and developments in \citet{Castelletti:2019}, a relative MCMC scheme on the EG space is constructed.
Our method is fully Bayesian and therefore outputs a posterior distribution over the space of essential graphs, rather than the single model estimate provided by the frequentist PC algorithm (\citet{Spir:Glym:Sche:2000}) available in the literaure. Accordingly, graph features of interest, such as the inclusion probabilities of specific edges, as well as measure of uncertainty around them, can be computed in our case. We will recover a single EG estimate for comparison purposes, and simulations will show that our method is competitive when adopted to recover a single EG structure. Furthemore, differently from the other Bayesian methods in the literature which implement the BDeu score of \citet{Heckerman:1995} on the space of DAGs, we directly score EGs by deriving a closed-form expression for the EG marginal likelihood and adopt an MCMC scheme targeting the posterior over the space of Markov equivalence classes. The benefits of a Bayesian method for DAG model selection specifically targeted to EGs rather than DAGs are addressed from a theoretical perspective by \citet{Ande:etal:1997}, and simulation comparisons will show that our approach is highly competitive under all scenarios, with outperformances in settings characterized by moderate sample sizes, especially with a high number of nodes.

The rest of the paper is organized as follows.
We first introduce some background material on DAGs and Bayesian analysis of categorical data in Section \ref{sec:background}.
The unstructured Bayesian inference of contingency tables outlined in Section \ref{sec:categorical:graphical:models} is extended to model selection of EGs in Section \ref{sec:graph:comparison}.
\black
Here we focus on the EG-driven likelihood decomposition, on the parameter prior induced by \citet{Geig:Heck:2002}, and on the derivation of the marginal likelihood with related asymptotic properties. The posterior sampler developed in Section \ref{sec:MCMC} is implemented on simulated data (Section \ref{sec:sim}), on a medical belief network and on US Congress voting records (Section \ref{sec:app}). We finally discuss extensions to intervential categorical data
and multiple datasets in Section \ref{sec:discussion}. 

\section{Background}
\label{sec:background}

In this section we provide some background material on Directed Acyclic Graphs (DAGs) and Essential Graphs (EGs) as well as on Bayesian analysis of contingency tables. Some futher notions on graphical models are reported in the Appendx. In addition, the reader can refer to \citet{Laur:1996} and the more recent book by \citet{roverato:2017} for a detailed exposition of these topics.
\black

\subsection{Directed Acyclic Graphs and Essential Graphs}


Let $\D=(V,E)$ be a DAG where $V=\{1,\dots,q\}$ is a set of nodes and $E\subseteq V\times V$ a set of directed edges 
and let $Y_1,\dots,Y_q$ be a collection of random variables that we associate to the nodes in $\D$.
A DAG encodes a set of conditional independencies between variables which defines its \textit{Markov property} and can be read off from the DAG using \textit{d-separation} (\citet{Pear:2000}).
Different DAGs may encode the same conditional independencies and accordingly we say that they are \textit{Markov equivalent}.
In many distributional settings, and in particular in the Gaussian framework and in the categorical case herein considered, Markov equivalent DAGs cannot be distinguished in the presence of only observational data; see also \citet{Geig:Heck:2002} and \citet{Heckerman:1995}.
Under further assumptions on the sampling distribution, such as equal variances within the Gaussian setting, Markov equivalence may not hold (\citet{Pete:Buhl:2014}) and DAGs can be in principle distinguished from observational data.

\citet[Theorem 1]{Verm:Pear:1990} shows that two DAGs are Markov equivalent if and only if they have the same \textit{skeleton} and the same \textit{v-structures}, therefore providing a graphical criterion to establish Markov equivalence.
For a given DAG $\D$, let $[\D]$ be its \textit{Markov equivalence class}, the set of all DAGs that are Markov equivalent to $\D$. By \citet{Ande:etal:1997} each equivalence class can be uniquely represented by a special \textit{chain graph} called \textit{essential graph} (EG), obtained as the union (over the edge sets) of Markov equivalent DAGs; an alternative name for an EG is \textit{completed partially directed acyclic graph} (CPDAG, \citet{Chic:2002}).
Finally, we recall an important result in \citet[Theorem 4.1]{Ande:etal:1997}, for which an EG is characterized as a chain graph with \textit{decomposable} chain components.


\subsection{Bayesian categorical data analysis}
\label{sec:categorical:graphical:models}

Let $Y_1, \dots, Y_q$ be specialized to a collection of categorical variables or \textit{classification criteria}, each $Y_j$ having set of levels $\Y_j$ and $l_j=|\Y_j|$.
We consider $n$ multivariate observations from $Y_1,\dots,Y_q$ where each $y^{i}$, $i=1,\dots,n$, corresponds to the levels of $Y_1,\dots,Y_q$ assigned to individual $i$, $y^i=\left(y^i(j), j\in V\right)$, and $y^i(j)$ denotes the $j$-th element in $y^i$, $V=\{1,\dots,q\}$.
These data can be collected into a $q$-dimensional \emph{contingency table} of counts $\bN$. To this end, let $\Y=\times_{j\in V}\Y_j$ be the product space generated by $Y_1,\dots,Y_q$, $y\in\Y$ an element of $\Y$, that is a generic configuration of the $q$ variables. Each count $n(y)$ representing the number of individuals assigned to configuration $y$ is given by
$
n(y) = \sum_{i=1}^{n}\mathbbm{1}(y^i=y),
$
being $\mathbbm{1}(\cdot)$ the indicator function. The collection of counts $n(y)$, $y\in\Y$, can be then arranged in a $q$-dimensional table.
Clearly $\sum_{y \in \Y}n(y)=n$ and the number of cells in $\bN$ coincides with the dimension of the product space $\Y$, that we denote by $l_V  = |\Y| = \prod_{j\in V}l_j$.
Let now $S\subseteq V$. A \emph{marginal table} of counts for the variables in $S$ is obtained by classifying the $n$ individuals \emph{only} according to criteria in $S$. The so-obtained $|S|$-dimensional marginal table is then $\bN_S$ with $l_S=\prod_{j\in S}l_j=|\Y_S|$ number of cells, where $\Y_S=\times_{j \in S}\Y_j$. For each cell $y_S \in \Y_S$ the corresponding count $n(y_S)$ is obtained from the original contingency table $\bN$ as $n(y_S) = \sum_{y\in\Y}n(y)\mathbbm{1}(y(S)=y_S)$
where $y(S)$ are the elements of $y \in \Y$ corresponding to variables in $S\in V$. 
It follows that $\sum_{y_S \in \Y_S}n(y_S)=n$.




For the generic cell $y \in \Y$, let $\theta_y$ the probability that an individual is assigned to configuration $y$, where $\sum_{y \in \Y}\theta_y = 1$.
The sampling distribution relative to an observation $y^i$ can be written as 
$
p(y^i\g \btheta) = \prod_{y\in\Y}\theta_y^{\mathbbm{1}(y^i=y)},
$
and the likelihood function for $n$ i.i.d. data points expressed as counts in the contingency table $\bN$ is then $p(\bN\g \btheta) = \prod_{y\in\Y}\theta_y^{n(y)}$,
where $\btheta = \{\theta_y,\ y\in\Y \}$ is the $l_V$-dimensional vector collecting the cell-probabilities $\theta_y$. 
For the methodology developed in the next sections we need a formula for the marginal distribution of the dataset $\bN$, namely
$
m(\bN) = \int p(\bN\g\btheta) \, p(\btheta) \, d\btheta
$,
where $p(\btheta)$ is a prior assigned to the model parameter $\btheta$.
A standard conjugate prior for $\btheta$ is the Dirichlet distribution,
$
\btheta \sim \textnormal{Dir}(\btheta \g \bA)\propto \prod_{y\in\Y} \theta_y^{\,a(y)-1}, 
$
where $a(y) \in \mathbb{R}^+$ and $\bA = (a(y), y\in\Y)$ denotes a $q$-dimensional table of hyperparameters with same size and structure of $\bN$. Because of conjugacy of the Dirichlet prior with model $p(\bN\g \btheta)$, the posterior 
$p(\btheta\g\bN)$ is $\textnormal{Dir}(\btheta\g \bA + \bN)$, where $\bA + \bN$ denotes the table collecting the element-by-element sums of $\bA$ and $\bN$. Accordingly, the marginal data distribution of $\bN$ can be obtained as the ratio of prior and posterior normalizing constants, so that
\ben
\label{eq:marg:like:complete}
m(\bN) =  \frac{\Gamma\left(\sum_{y \in \Y} a(y) \right)}{\Gamma\left(\sum_{y \in \Y} (a(y) + n(y)) \right)} \,
\prod_{y \in \Y} \, 
\frac{\Gamma(a(y)+n(y))}{\Gamma(a(y))}.
\een
\normalsize
Different choices for the hyperparameters of the Dirichlet prior are possible.
If for simplicity we set $a(y) = a$, Equation \eqref{eq:marg:like:complete} reduces to $\Gamma\big(l_Va \big)/\Gamma\big(l_Va + n\big)\prod_{y \in \Y}\{\Gamma(a+n(y))/\Gamma(a)\}.$

Consider now a subset $S\subseteq V$, with implied marginal table $\bN_S$.
For later developments, we also need a formula for the marginal data distribution of $\bN_S$.
Recall that $n(y_S)$ is the count corresponding to the cell $y_S\in\Y_S$ appearing in $\bN_S$.
The likelihood function restricted to $\bN_S$ can be written as $p(\bN_S\g\btheta_S) = \prod_{y_S \in \Y_S} \theta_{y_S}^{\,n(y_S)}$,
where
$
\theta_{y_S}=\sum_{y \in \Y}\theta_{y} \mathbbm{1}(y(S)=y_S)
$
are the marginal probabilities for variables in $S$ and $\btheta_S$ is the vector  of dimension $l_S$ collecting the cell-probabilities $\theta_{y_S}$. 
Moreover, for the aggregation property of the Dirichlet distribution we have
$
\btheta_S \sim \textnormal{Dir}(\btheta_S \g \bA_S),
$
where $\bA_S=(a(y_S), y_S\in\Y_S)$ is a $|S|$-dimensional table of hyperparameters with elements $a(y_S)$ given by
$
a(y_S)=\sum_{y \in \Y}a(y) \mathbbm{1}(y(S)=y_S).
$
Accordingly, the posterior distribution of $\btheta_S$ is $\textnormal{Dir}(\btheta_S\g \bA_S + \bN_S)$ and the marginal data  distribution restricted to the table $\bN_S$ is
\ben
\label{marg:like:complete:sub}
m(\bN_S) = \frac{\Gamma\left(\sum_{y_S\in\Y_S} a(y_S) \right)}{\Gamma\left(\sum_{y_S\in\Y_S} (a(y_S) + n(y_S)) \right)} \, \prod_{y_S\in\Y_S}
\frac{\Gamma(a(y_S)+n(y_S))}{\Gamma(a(y_S))}.
\een
\normalsize
Note that, if we let $a(y) = a$, we obtain, with $\bar{S}=V\setminus S$,
$
a(y_S) = a\sum_{y\in\Y}\mathbbm{1}(y(S)=y_S) = l_{\bar{S}} \, a.
$

\section{Model comparison of essential graphs}
\label{sec:graph:comparison}



\black

In this section we instead focus on EGs and derive a closed-form expression for the marginal likelihood of a categorical EG model.
We first write the likelihood function which factorizes according to the graphical structure imposed by the EG (Section \ref{sec:EG:likelihood}).
The latter involves a collection of parameters for each chain component (decomposable UG) for which a suitable prior distribution must be specified.
To this end we follow the procedure of \citet{Geig:Heck:2002} originally introduced for model comparison of DAG models.
\black
The EG marginal likelihood 
is obtained in Section \ref{sec:EG:marginal:likelihood}, with related asymptotic properties studied in Section \ref{sec:asym}, and considerations on the hyperparameter choice in Section
\ref{sec:hype}.

\black
\subsection{Likelihood decomposition}
\label{sec:EG:likelihood}
Let $\G=(V,E)$ be an EG. Recall from \citet[Theorem 4.1]{Ande:etal:1997} that $\G$ is a chain graph where each chain component $\tau\in\T$, $\tau \subseteq V$, corresponds to a decomposable UG $\G_{\tau}$.
Let also $y \in \Y$ and $y_{\tau}\in\Y_{\tau}$ be the generic element of the product spaces $\Y$ and $\Y_{\tau}$ respectively as defined in Section \ref{sec:categorical:graphical:models}; similarly for $y_{\pa_\G(\tau)}\in \Y_{\pa_\G(\tau)}$, where $\pa_{\G}(\tau)$ denotes the set of parents of $\tau$ in $\G$.
For simplicity of notation we will omit the subscript $\G$ (e.g. by writing $\pa(\tau)$ instead of $\pa_\G(\tau)$) so that the dependence on the underlying EG will be tacitly assumed. All the results presented below are therefore predicated on a given EG $\G$.

Recall first from \citet{Ande:etal:2001} that under a given EG the probability distribution related to an observation $y \in \Y$ factorizes as
\ben
\label{eq:EG:fact:density:2}
p(y\g \btheta) = \prod_{\tau\in\T}p\big(y(\tau)\g y(\pa(\tau)),\btheta_{\tau\g y(\pa(\tau))}\big),
\een
where $\btheta$ is a global parameter indexing the EG model, while 
\black
$\btheta_{\tau\g y(\pa(\tau))} = \{\theta_{y_\tau|y(\pa(\tau))},\ y_\tau \in \Y_\tau\}$ is a local parameter for chain component $\tau$, 
corresponding to configurations of variables in $\pa(\tau)$ actually \emph{observed}; 
see also \citet[Equation 3]{Cast:Perl:2004}. Accordingly, the likelihood function  for a complete dataset $\bD$ comprising $n$ observations $y^i$, $i=1,\dots, n$, can be written as
\ben
p(\bD\g \btheta) &=& 
\prod_{i=1}^{n} \prod_{\tau\in\T} p\left(y^i(\tau) \,\big|\, y^i(\pa(\tau)),\btheta_{\tau\g y^i(\pa(\tau))}\right)\nonumber\\
&=& \prod_{i=1}^{n}
\prod_{\tau\in\T}
\prod_{r\in\Y_{\pa(\tau)}}
\prod_{s\in\Y_\tau}
\theta_{s\g r}^{\mathbbm{1}\left\{y^i(\tau) = s, \, y^i(\pa(\tau)) = r\right\}}
, \nonumber
\een
by expanding the product over the sets $\Y_{\pa(\tau)}$ and $\Y_{\tau}$.
Therefore 
\ben
\label{eq:EG:fact:likelihood:bis}
p(\bN\g\btheta)
&=&
\prod_{\tau\in\T}
\prod_{r\in\Y_{\pa(\tau)}}
\prod_{s\in\Y_\tau}
\theta_{s\g r}^{n(s \g r)}
\nonumber \\
&=&
\prod_{\tau\in\T}
\prod_{r\in\Y_{\pa(\tau)}}
p(\bN_{\tau}\g \bN_{\pa(\tau)}, r, \btheta_{\tau\g r}),
\een
where the conditional frequency $n(s \g r)$ corresponds to the number of observations assigned to level $s$ and $r$ of variables in $\tau$ and $\pa(\tau)$ respectively. 
Equation \eqref{eq:EG:fact:likelihood:bis} corresponds to the likelihood for $n$ i.i.d. observations expressed as counts in the contingency table $\bN$ respecting the graphical structure imposed by the EG.


\subsection{Parameter prior distributions}
\label{sec:EG:parameter:prior}

\citet{Heckerman:1995} and \citet{Geig:Heck:2002} (G\&H) propose a method for the construction of parameter priors on DAG models. An important implication of their approach concerns the computation of the marginal likelihood of any DAG, which can be directly obtained from the marginal data distribution computed under a complete model.
In more details, \citet{Heckerman:1995} introduce an elicitation procedure for prior parameter construction across DAG models and decomposable UG models. 
Starting from few assumptions that are naturally satisfied in the Gaussian setting by Normal-Wishart priors and in the categorical framework by Dirichlet priors (\citet{Geig:Heck:2002}),
they show how to assign a prior to the parameters of any given DAG (or decomposable UG) starting from a unique prior assigned to the parameter of a \textit{complete} DAG (or decomposable UG) model. 
In our EG context, we implement this elicitation procedure at the level of chain component, since each chain component corresponds to a decomposable UG (Theorem 4.1 of \citealt{Ande:etal:1997}).
This approach dramatically simplifies the prior elicitation procedure across EGs and provides a default method to assign priors to EG model parameters:
we are then allowed to assume standard Dirichlet priors, in accordance to Section \ref{sec:categorical:graphical:models}.

For the EG global parameter $\btheta$ we first assume that the prior factorizes as
\begin{eqnarray}
\label{prior:indep:tau}
p(\btheta)=\prod_{\tau \in {\mathcal{T}}} p(\btheta_{\tau}),
\end{eqnarray}
a condition known as \textit{global independence}, which extends the assumption of global parameter independence, typical of DAG models, to CG models; see also \citet{Cast:Perl:2004}.
For any $\tau\in\T$ consider now $\btheta_{\tau\g r}=\{\theta_{y_\tau\g r}, y_\tau\in\Y_{\tau}\}$, $r\in\Y_{\pa(\tau)}$. We further assume \textit{local independence}, namely that $\btheta_{\tau\g r}$ are \textit{a priori} independent:
\begin{eqnarray}
\label{prior:indep:tau:r}
p(\btheta_{\tau})=\prod_{r\in\Y_{\pa(\tau)}} p(\btheta_{\tau\g r}).
\end{eqnarray}

Recall that each $\btheta_{\tau\g r}$ consists of a
vector of (conditional) probabilities $\theta_{y_{\tau}\g r}$, $y_{\tau} \in \Y_{\tau}$. 
Assuming that the underlying (decomposable) sub-graph $\G_{\tau}$ is complete, we can set
\ben
\label{eq:prior:dir:complete}
p(\btheta_{\tau\g r}) = \textnormal{pDir}(\btheta_{\tau\g r} \g \bA_{\tau \g r}) 
\propto \prod_{y_{\tau}\in\Y_{\tau}} \theta_{y_{\tau}\g r}^{\,a(y_{\tau}\g r)-1}. \label{eq:prior:tau}
\een

Let now $S\subseteq \tau$
and $\bN_{S}$ be the corresponding marginal table; see also Section \ref{sec:categorical:graphical:models}.
Accordingly, the likelihood function restricted to $S$ can be written as
\ben
p(\bN_{S}\g \bN_{\pa(\tau)}, r, \btheta_{S\g r}) = \prod_{s \in \Y_S} \theta_{s\g r}^{\,n(s\g r)} 
,
\een
where 
\black
$
\theta_{s\g r}=\sum_{y_{\tau} \in \Y_{\tau}}\theta_{y_{\tau}\g r} \mathbbm{1}(y_{\tau}(S)=s),
$
whilst, fixing $a(s\g r)=\sum_{y_{\tau} \in \Y_{\tau}}a(y_{\tau}\g r) \mathbbm{1}(y_{\tau}(S)=s)$, 
\ben
&& p(\btheta_{S\g r}) = \textnormal{pDir}(\btheta_{S\g r} \g \bA_{S \g r})
\propto \prod_{s\in\Y_S} \theta_{s\g r}^{\,a(s \g r)-1}
\een
is the prior induced by \eqref{eq:prior:tau}.

\subsection{Marginal likelihood of EG models}
\label{sec:EG:marginal:likelihood}

We now focus on the computation of the marginal likelihood of $\G$,
\ben
\label{marg:like:EG}
m_{\G}(\bN)
= \int p_{\G}(\bN \g \btheta_{\G})p(\btheta_{\G})\,d\btheta_{\G},
\een
where we now emphasize the dependence on the EG $\G$.
Because of the independence assumptions in \eqref{prior:indep:tau}, we can write
\begin{eqnarray}
\label{eq:marg:like:fact}
m_{\G}(\bN)
&=& \prod_{\tau \in \T}
\,
\int p_{\tau}(\bN_{\tau} \g \bN_{\pa_{\G}(\tau)}, \btheta_{\tau})
p(\btheta_{\tau})\,d\btheta_{\tau} \nonumber \\ 
&=& \prod_{\tau \in \T}m_{\tau}(\bN_{\tau} \g \bN_{ \pa_{\G}(\tau)}).
\end{eqnarray}
where it appears that $m_{\G}(\bN)$ admits the same factorization of the sampling density in \eqref{eq:EG:fact:likelihood:bis}.
Next, 
because of the independence assumption across $\btheta_{\tau\g r}$ in \eqref{prior:indep:tau:r}, we can write
\be
m_{\tau}(\bN_{\tau} \g \bN_{ \pa_{\G}(\tau)})
=
\prod_{r\in\Y_{\pa_{\G}(\tau)}}
m_{\tau\g r}(\bN_{\tau}\g \bN_{\pa_{\G}(\tau)}, r).
\ee
Recall that for a decomposable UG $\G_{\tau}$ with sets of cliques and separators $\mathcal{C}_{\Gtau}$ and $\mathcal{S}_{\Gtau}$, the marginal likelihood $m_{\tau|r}(\cdot)$ admits the factorization of \citep{Laur:1996}:

\small
\ben
\label{like:decomposable:chain}
m_{\tau \g r}(\bN_{\tau}\g \bN_{\pa_{\G}(\tau)}, r) =
\frac{\prod_{C\in \mathcal{C}_{\tau}}
	m(\bN_C \g \bN_{ \pa_{\G}(\tau)},r)}
{\prod_{S\in \mathcal{S}_{\tau}}
	m(\bN_S \g \bN_{ \pa_{\G}(\tau)},r)}.
\een
\normalsize
In addition, because of the theory presented in \citet{Geig:Heck:2002} and applied to decomposable UGs by \citet{Cons:Laro:2012}, each term $m(\bN_S \g \cdot)$ in \eqref{like:decomposable:chain} corresponds to the marginal data distribution computed under a complete graph as in
Equation \eqref{marg:like:complete:sub}, 
and similarly for $m(\bN_C \g \bN_{ \pa_{\G}(\tau)},r)$, that is

\small
\ben
m(\bN_S \g \bN_{ \pa_{\G}(\tau)},r) =
\frac{\Gamma\left(\sum_{s\in\Y_S} a(s \g r) \right)}
{\Gamma\left(\sum_{s\in\Y_S} (a(s \g r) + n(s \g r)) \right)}
\cdot \prod_{s\in\Y_S}
\frac{\Gamma\big(a(s \g r )+n(s \g r)\big)}
{\Gamma\big(a(s \g r)\big)}.
\een
\normalsize
\black
Note that the total number of parameters is 
$|\btheta_\G|=\sum_{\tau}l_{\fa(\tau)}$, where $\fa(\tau) = \pa(\tau)\cup\tau$.
We stress that we can handle the high-dimensional case of $n<<|\btheta_\G|$, with no constraints on the sparsity of the graph, differently from the Gaussian context of \citet{Cons:Laro:2012} and \citet{cons:et:al:2017}, where the minimum number of observations is related to the \textit{clique number} of the graph, the dimension of the largest maximal clique.

\subsection{Asymptotic behaviour of the marginal likelihood}
\label{sec:asym}
In this section we derive the asymptotic distribution of the marginal likelihood. More precisely we show, for a single clique or separator of the graph, that the logarithm of the marginal likelihood, scaled by a factor of $\sqrt{n}$, converges in distribution to a Gaussian random variable when the number of observations diverges, conditionally to the knowledge of parents configurations. The asymptotic variance, for which we provide an easy estimator, reveals the speed of convergence at which the marginal likelihood converges to its asymptotic mean, that is to the marginal likelihood evaluated at the true population configuration probabilities. 

We first fix $\tilde n(c|r)=n(c|r)/n(r)$ as the observed relative frequency of a configuration $c \in \Y_C$, given that $y_{\pa(C)} = r$; similarly for $\tilde n(s|r)$, $s \in \Y_S$. The hyperparameters $a(c|r)$ and $a(s|r)$ are implied by $a(y_\tau|r)$, $y_\tau \in \Y_\tau$, through the aggregation property of the Dirichlet distribution. 
Given a generic separator $S \in \tau$ (but the same can be stated for a clique $C$), it is a standard result that 
\ben
\{ \tilde n(s|r), s \in \Y_S \} \xrightarrow{d} \N_{l_S}\left(\theta^0_{S\g r},\Sigma_{S|r}/n(r)\right), \label{eq:normal}
\een
where $\theta^0_{S\g r}$ is the vector of the true configuration probabilities in $S$, given a specific configuration $r$ of the parents, and where $\Sigma_{S|r} = (\sigma_{s_1,s_2|r})_{l_S \times l_S}$, with
\ben
\sigma_{s_1,s_2|r} = \left\{\begin{array}{ll}
\theta^0_{s_1|r}(1-\theta^0_{s_1|r}), & s_1=s_2\\
-\theta^0_{s_1|r}\theta^0_{s_2|r} , & s_1\neq s_2\nonumber
\end{array}\right..
\een
If $a(s|r)$ is chosen so that $a(s|r)/n(r) \to 0$, the result in \eqref{eq:normal} is also valid for $\{ \bar n(s|r), y_\tau \in \Y_S \}$, where $\bar n(s|r):=(n(s|r)+a(s|r))/n(r)$, for instance when $a(s|r)=a_S$ is a constant depending on the set $S$ but not on the configurations of nodes in $S$ or parents. Furthermore, assuming for simplicity $a(s|r)=a$, we have that
\ben
\frac1{n(r)}\log m(\bN_S \g \bN_{ \pa(\tau)},r) &=&
C_1(a,S,r) + \frac1{n(r)} \sum_{s\in\Y_S}\log\Gamma\left(n(r)\bar n(s|r)\right) \nonumber\\
&-&\frac1{n(r)} \log\Gamma\left(n(r) \sum_{s\in\Y_S} \bar n(s|r)\right) \nonumber\\
&=:& g_{S|r}\big(\{\bar n(s|r),\ s\in \Y_S\}\big)
\een
\normalsize
for some function $C_1$ depending on $a$, $S$ and $r$, but not on the data, such that $C_1(a,S,r)\to0$ as $n(r)\to\infty$. Since $g_{S|r}$ is continuous and with at least one non-null partial derivative in $\bar n(s|r)$ fixed to $\theta_{s\g r}^0$, all $s \in \Y_S$, by the Delta method the asymptotic normality is preserved, with 
\ben
\frac1{n(r)}\log m(\bN_S \g \bN_{ \pa(\tau)},r) \xrightarrow{d}
\N\left( g_{S|r}\left(\{\theta^0_{s|r},\ s\in \Y_S\}\right),D_{S|r}'\Sigma_{S|r}D_{S|r}/n(r) \right),\nonumber
\een
where $D_{S|r}=\{D_{s|r},\ s\in \Y_S\}$, $D_{s|r}= n(r)\psi(n\theta_{s\g r}^0)-n(r)\psi\left(n(r)\right)$ and $\psi$ is the digamma function. From the approximation $\exp(\psi(x))\approx x-1/2$, valid for large $x$, we can write, for $n(r)$ large enough,
$$
\psi\left(n(r)\theta_{s\g r}^0\right) \approx \log\left( n(r)\theta_{s\g r}^0 - 1/2 \right)
$$
and
$$
D_{s|r} \approx \log\left( \frac{n(r)\theta_{s\g r}^0-1/2}{n(r)-1/2} \right) \approx \log\theta_{s\g r}^0,
$$
so that the asymptotic variance becomes
\ben
D_{S|r}'\Sigma_{S|r}D_{S|r}/n(r) \approx
\frac1{n(r)}\sum_{s_1\in\Y_S}\sum_{s_2\in\Y_S} \sigma_{s_1,s_2|r} \log\theta_{s_1\g r}^0\log\theta_{s_2\g r}^0\nonumber
\een
and then finally
\ben
\frac{\sqrt{n(r)}\left(g_{S|r}\left(\{\bar n(s|r),\ s\in \Y_S\}\right) - g_S\left(\{\theta^0_{s|r},\ s\in \Y_S\}\right) \right)}{\left(\sum_{s_1\in\Y_S}\sum_{s_2\in\Y_S} \sigma_{s_1,s_2|r} \log\theta_{s_1\g r}^0\log\theta_{s_2\g r}^0\right)^{1/2}}
\xrightarrow{d} \N(0,1). \label{eq:normal_g}
\een
\normalsize
From convergence in probability of $ \bar n(s|r)$ and continuous mapping theorem, the result in \eqref{eq:normal_g} is also valid with the denominator replaced by its estimate
$$
\left(\sum_{s_1\in\Y_S}\sum_{s_2\in\Y_S} \hat\sigma_{s_1,s_2|r} \log \bar n(s_1| r)\log \bar n(s_2|r)\right)^{1/2},
$$
where 
$$
\hat\sigma_{s_1,s_2|r} = \left\{\begin{array}{ll}
\bar n(s_1|r)(1-\bar n(s_1|r)), & s_1=s_2\\
-\bar n(s_1|r)\bar n(s_2|r) , & s_1\neq s_2
\end{array}\right..
$$
See also that if we replace $n(r)$ by $n$, we can repeat the steps above for 
$$
\bar g_{S|r}\left(\{\bar n(s|r),\ s\in \Y_S\}\right):=\frac1{n}\log m(\bN_S \g \bN_{ \pa(\tau)},r)
$$
and obtain, asymptotically in $n$, that
\ben
\frac{\sqrt{n}\left(\bar g_{S|r}\left(\{\bar n(s|r),\ s\in \Y_S\}\right) - \bar g_S\left(\{\theta^0_{s|r},\ s\in \Y_S\}\right) \right)}{\sqrt{\theta^0_r\sum_{s_1\in\Y_S}\sum_{s_2\in\Y_S} \sigma_{s_1,s_2|r} \log\theta_{s_1\g r}^0\log\theta_{s_2\g r}^0}}
\xrightarrow{d} \N(0,1),
\een
still valid with $\hat\sigma_{s_1,s_2|r}$, $\bar n(s|r)$ and $n(r)/n$ replacing, respectively, $\sigma_{s_1,s_2|r}$, $\theta_{s\g r}^0$ and $\theta^0_r$ in the denominator.
Therefore with $a(s|r)/n \to 0$, an appropriately scaled version of $\log m(\bN_S \g \bN_{ \pa(\tau)},r)$ is asymptotically Gaussian and correctly centered, with a variance that can be estimated.

\subsection{On the hyperparameter choice}\label{sec:hype}
Following \citet{Geig:Heck:2002}, we start from a unique prior at level of chain component $\tau$ and all other priors for included cliques and separators are derived accordingly, in a way that is coherent with the hyperparameter construction in the BDeu score of \citet{Heckerman:1995}. As pointed out in \citet{scutari2018}, BDeu is the only score that guarantees equal scores to Markov equivalent DAGs; see also \citet{scutari2016empirical}.
Still, we stress that our marginal likelihood derived in Section \ref{sec:EG:marginal:likelihood} does not coincide with the BDeu score of \citet{Heckerman:1995}, since the latter is on DAGs and not on EGs. Only the part of our marginal likelihood related to a single chain component and conditionally to one observed configuration of the parent nodes can be reconducted to the BDeu form.

Any possible value for $a(s|r)$ for which $a(s|r)/n \to 0$ guarantees the validity of the results in Section \ref{sec:asym}. By choosing for all $y_\tau \in \Y_\tau$ and $r\in\Y_{\pa(\tau)}$, $a(y_\tau|r)<1$,  
we opt for the sensible choice of a prior distribution with no mode on any chain configuration. We want this property to be valid also for all clique and separator configurations within the chain component. Furthermore, a prior choice of $\mathbb{V}(\theta_{y_\tau|r}) = \alpha$ implies $\mathbb{V}(\theta_{c|r}) \approx \alpha l_\tau/l_C$, meaning more prior uncertainty for probabilities associated to smaller cliques or separators, proportionally to their dimension, relative to the dimension of the chain component they belong. Then, to have the same prior information on cliques/separators of same dimension in different chain components, to impose no prior mode on any cliques/separators configurations, and for results in Section \ref{sec:asym} to be valid, we ultimately suggest the choice of $a(y_\tau|r) = 1/l_\tau$.

\section{Computational implementation}
\label{sec:MCMC}

In this Section we introduce the MCMC scheme that we adopt to sample from the posterior distribution on the EG space and perform posterior model inference of categorical EGs.

\subsection{MCMC scheme}
\label{sec:MCMC:scheme}

Let $\mathcal{S}_q$ be the set of all EGs on $q$ nodes. Our MCMC consists of a Metropolis Hastings (MH) algorithm targeting the posterior distribution on the EG space,
\be
p(\G \g \bN) \propto m_{\G}(\bN) \, p(\G), \quad \G \in \mathcal{S}_q,
\ee 
where $m_{\G}(\bN)$ is the marginal likelihood of $\G$ computed as in Equation \eqref{eq:marg:like:fact}, $p(\G)$ a prior assigned to $\G$.
A similar scheme was introduced in \citet{Caste:etal:2018} within the context of Gaussian EGs. The key feature of this algorithm is the choice of a suitable proposal distribution which determines the transitions between EGs belonging to the (discrete) model space $\mathcal{S}_q$.
To this end, \citet{Caste:etal:2018} adopted the Markov chain originally proposed by \citet{He:etal:2013} to explore the EG space and investigate features of interest (such as the number of directed and undirected edges, \textit{v}-structures and so on). Some optimality properties, namely irreducibility and reversibility, allows to efficiently compute the stationary distribution of the Markov chain, used to weigh samples obtained from the proposal distribution.

Transitions between EGs are determined by six types of operators: inserting an undirected edge (denoted by InsertU), deleting an undirected edge (DeleteU), inserting a directed edge (InsertD), deleting a directed edge (DeleteD), converting two adjacent undirected edges in a \textit{v}-structure (MakeV) and converting a \textit{v}-structure into two adjacent undirected edges (RemoveV). Besides these, following \citet{Caste:Cons:2019} we also adopt the operator ReverseD originally introduced by \citet{Chic:2002}. 
Such operator is not needed for the Markov chain to be irreducible and reversible, but it adds extra-connectivity to the states of the chain, thus improving the exploration of the EG space; see also \citet{Castelletti:2019} for a general presentation of the MCMC scheme.
For each EG $\G$ we can then construct a set of \textit{perfect} operators $\mathcal{O}_{\G}$, i.e. guaranteeing that the resulting graph is an EG. Let $\mathcal{O}_{\G}$ be a perfect set of operators on $\G$, $|\mathcal{O}_{\G}|$ its cardinality. It can be shown that the probability of transition from $\G$ to $\G'$, the latter a direct successor of $\G$, is
$$
q(\G'\g\G) = 1/|\mathcal{O}_{\G}|.
$$

Next, we need to specify a prior $p(\G)$, for $\G \in \mathcal{S}_q$.
Let $\bA^{\G}$ be the (symmetric) 0-1 adjacency matrix of the skeleton of $\G$, whose $(u,v)$ element is denoted by $\bA^{\G}_{(u,v)}$.
Conditionally on a probability of edge inclusion $\pi\in (0,1)$,
we first assign a Bernoulli prior independently to each element $\bA^{\G}_{(u,v)}$ 
in the lower triangular part of $\bA^{\G}$,
$\bA^{\G}_{(u,v)} \g \pi \stackrel{iid}{\sim} \textnormal{Ber}(\pi), u > v$.
Therefore,
\ben
\label{eq:prior:skeleton}
p(\bA^{\G})=
\pi^{|\bA^{\G}|}(1-\pi)^{\frac{q(q-1)}{2}-|\bA^{\G}|},
\een
where $|\bA^{\G}|$ denotes the number of edges in the skeleton of $\G$ and $q(q-1)/2$ corresponds to the maximum number of edges in the graph.
We finally set $p(\G)\propto p(\bA^{\G})$ for each $\G \in \mathcal{S}_q$,
which results in a simple prior only depending on the number of edges in the graph and that can easily reflect prior knowledge of sparsity (\citet{Caste:etal:2018}). Other priors, specific for DAGs and based on the number of compatible perfect orderings of the vertices, are also present in the literature (\citet{Friedman:Koller:2003,Kuipers:Moffa:2017}).

\black
Let $m_{\G}(\bN)$ be the marginal likelihood of $\G$ given the table of counts $\bN$, $p(\G)$ a prior on $\G$ and $q(\G'\g\G)$ a proposal distribution for the chain when we are at graph $\G$. At each step of the MH scheme we then propose a new EG $\G'$ given the current graph $\G$ from $q(\G'\g \G)$ and accept $\G'$ with probability
\begin{equation}
\label{eq:acceptancealpha}
\alpha_{\G,\G'}=\min\left\{ 1; \frac{m_{\G'}(\bN)}{m_{\G}(\bN)}\cdot \frac{p(\G')}{p(\G)} \cdot
\frac{q(\G\g\G')}{q(\G'\g\G)}\right\}.
\end{equation}

\subsection{Posterior model inference}
\label{sec:MCMC:posterior_model:inference}

Our MCMC output consists of a collection of EGs visited by the chain, $\{\G^{(1)},\dots,\G^{(T)}\}$. This can be used to approximate the posterior distribution over the EG space as
\ben
	p(\G \g \bN)
	=\frac{m_{\G}(\bN)p(\G)}{\sum_{\G\in\mathcal{S}_q} m_{\G}(\bN)p(\G)}
	\approx \frac{1}{T}\sum_{t=1}^{T} \mathbbm{1}\left\{\G^{(t)}=\G\right\},\nonumber
\een
where $\mathbbm{1}(\cdot)$ is the indicator function; 
see also \citet{Garcia:Donato:et:al:2013} for a discussion on frequency-based estimators in large model spaces.
In addition we can recover from the same output the (estimated) posterior probability of inclusion for each (directed) edge,
\ben
\label{eq:posterior:edge:inclusion}
\hat{p}_{u\rightarrow v}(\bN)
=\frac{1}{T}\sum_{t=1}^{T}\mathbbm{1}_{u\rightarrow v}\left\{\G^{(t)}\right\},
\een
where $\mathbbm{1}_{u\rightarrow v}\{\G^{(t)}\}=1$ if $\G^{(t)}$ contains $u\rightarrow v$, 0 otherwise, and an undirected edge $u-v$ is equivalent to the union of $u \rightarrow v$ and $u \leftarrow v$.
Starting from these quantities a single EG estimate summarizing the whole output, if required, can be also obtained.
For instance, one can consider the \textit{maximum a posteriori (graph) model} (MAP) which corresponds to the EG with highest associated posterior probability. However, the MAP may not represent an optimal choice especially from a predictive viewpoint as discussed for instance by \citet{Barb:Berg:2004} in a multiple linear regression framework. Differently, it was shown that the \textit{median probability model}, which in their context was obtained by including all variables whose posterior probability of inclusion exceeds 0.5, is predictively optimal.
In our EG setting, we can proceed similarly and construct first a graph estimate (that we name median probability \textit{graph} model) by including all edges $u \rightarrow v$ such that $\hat{p}_{u \rightarrow v}(\bN)>0.5$. Since the latter is not guaranteed to be an EG, while is in general a partially directed graph, one further possibility is to consider any consistent extension \citep{Dor:Tars:1992} of the median probability model, as detailed in \citet{Caste:etal:2018}. The resulting EG estimate is called \textit{projected} median probability graph model.

\section{Simulations}\label{sec:sim}

We now evaluate the performance of our method through simulations.
Specifically, we vary the number of variables $q\in\{5,10,20,40\}$ and the sample size $n\in\{100,$ $200,500,1000\}$. For each combination of $q$ and $n$ (a scenario) we generate 40 categorical datasets as detailed in Section \ref{subsec:data:generation}. For simplicity we assume all variables being binary, namely $Y_j\in\{0,1\}$, $j=1,\dots, q$. Results and comparisons with some benchmark methods are presented in Section \ref{sub:sec:simulation:results}.

\subsection{Data generation}
\label{subsec:data:generation}

For a given value of $q$ we first randomly generate $40$ DAGs using the function \texttt{randomDAG} in the R package \texttt{pcalg} by fixing a probability of edge inclusion equal to $p_{edge}=3/(2q-2)$ as in the sparse setting of \cite{Pete:Buhl:2014}. Each DAG $\D$ defines a data generating process which in a Gaussian setting \citep{Caste:etal:2018} we can write as
\ben
\label{eq:generating:process}
Z_{i,j} = \mu_j + \sum_{k\in \pa_\D(j)} \beta_{k,j}Z_{i,k} + \varepsilon_{i,j},
\een
for $i=1,\dots,n$ and $j=1,\dots,q$, where $\varepsilon_{i,j}\sim \N(0,\sigma_j^2)$ independently.
For each $j$ we fix $\mu_j = 0$ and $\sigma^2_j=1$, while regression coefficients $\beta_{k,j}$ are uniformly chosen in the interval $[-1,-0.1] \cup [0.1,1]$; see also \citet{Pete:Buhl:2014}.
For expediency we then proceed by generating first $n$ multivariate Gaussian observations from \eqref{eq:generating:process}; a categorical dataset consisting of $n$ observations from $q$ binary variables is then obtained by setting
\begin{equation}
\label{eq:discretize:gaussian}
Y_{i,j}=
\begin{cases}
\,\, 1 & \text{if } Z_{i,j}\geq \gamma_j, \\
\,\, 0 & \text{if } Z_{i,j}< \gamma_j,
\end{cases}
\end{equation}
where we fix $\gamma_j=0$, for $j=1,\dots,q$.
Finally, for each DAG $\D$ we consider its representative EG
which will represent the benchmark of comparison with the EG estimate provided by each method under evaluation; more details are given in the next section.

\subsection{Simulation results}
\label{sub:sec:simulation:results}

We evaluate the performance of our method, that we name DBEG (Discrete Bayesian EG), in recovering the graphical structure of the true EG.
To this end, for each $q\in\{5,10,20,40\}$ we run $T=1000 \cdot q$ iterations of our MCMC algorithm (Section \ref{sec:MCMC}). To favour sparsity, we fix the hyperparameter $\pi$ in the EG prior \eqref{eq:prior:skeleton} as $\pi=1.5/(2q-2)$ which corresponds to a prior probability of edge inclusion smaller than the expected level of sparsity, as commonly recommended; see for instance \citet{Pete:etal:2015}. Finally, we fix  $a(y_{\tau}\g r)=1/l_{\tau}$ in the Dirichlet prior \eqref{eq:prior:dir:complete}, as suggested in Section \ref{sec:hype}.

We compare our method with the PC algorithm for categorical data of \citet{Spir:Glym:Sche:2000}, a constraint-based method that estimates the EG through multiple conditional independence tests, at a significance level $\alpha$ that we fix as $\alpha\in\{0.10,0.05,0.01\}$. As other benchmarks, we use HC Bdeu, an optimized hill climbing greedy search that explores the space of DAGs by single-arc additions, removals and reversals and that uses the BDeu score of \citet{Heckerman:1995}, and TABU BDeu (\citet{russell2002artificial}), a modified hill-climbing algorithm able to escape local optima by selecting DAGs that minimally decrease the score function. Since both HC BDeu and TABU BDeu were not specifically designed for EGs but for DAG model selection, their DAG estimates are converted in the EG representative of the corresponding equivalence class.

\begin{figure*}
		\begin{center}
			\includegraphics[scale=0.2]{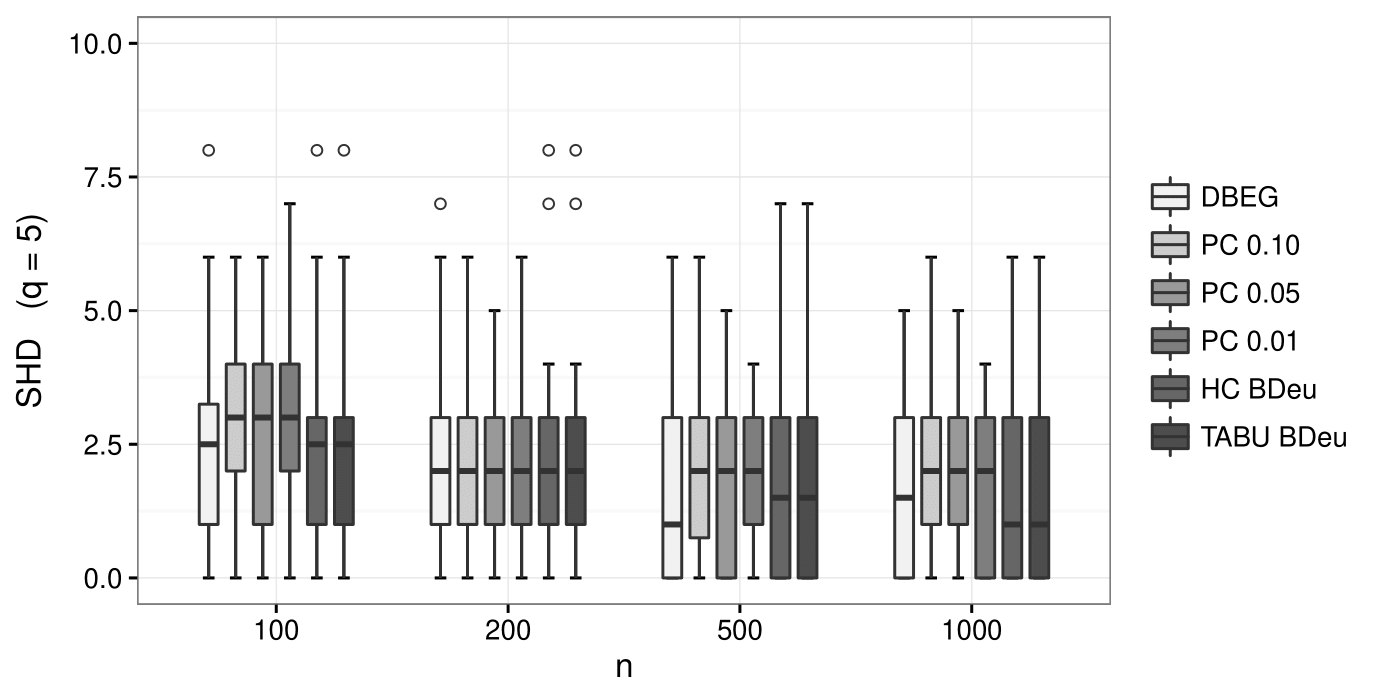}
			\includegraphics[scale=0.2]{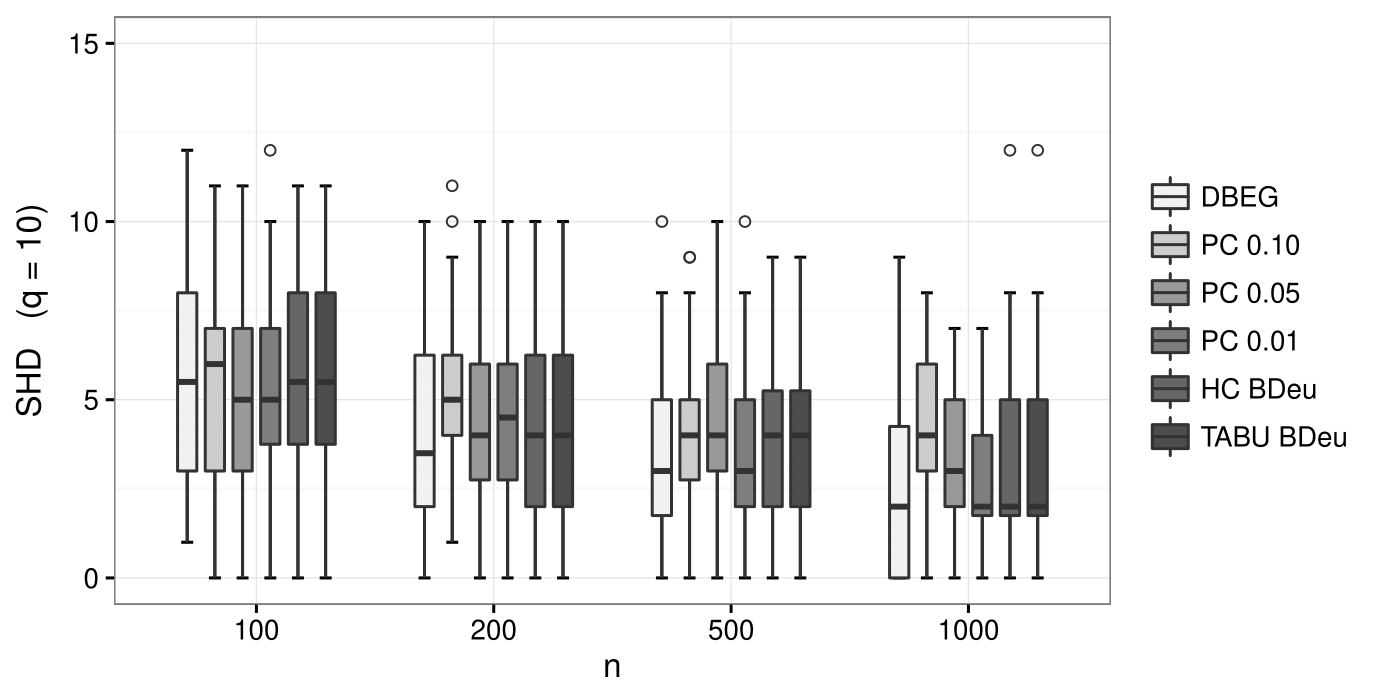} \\
			\includegraphics[scale=0.2]{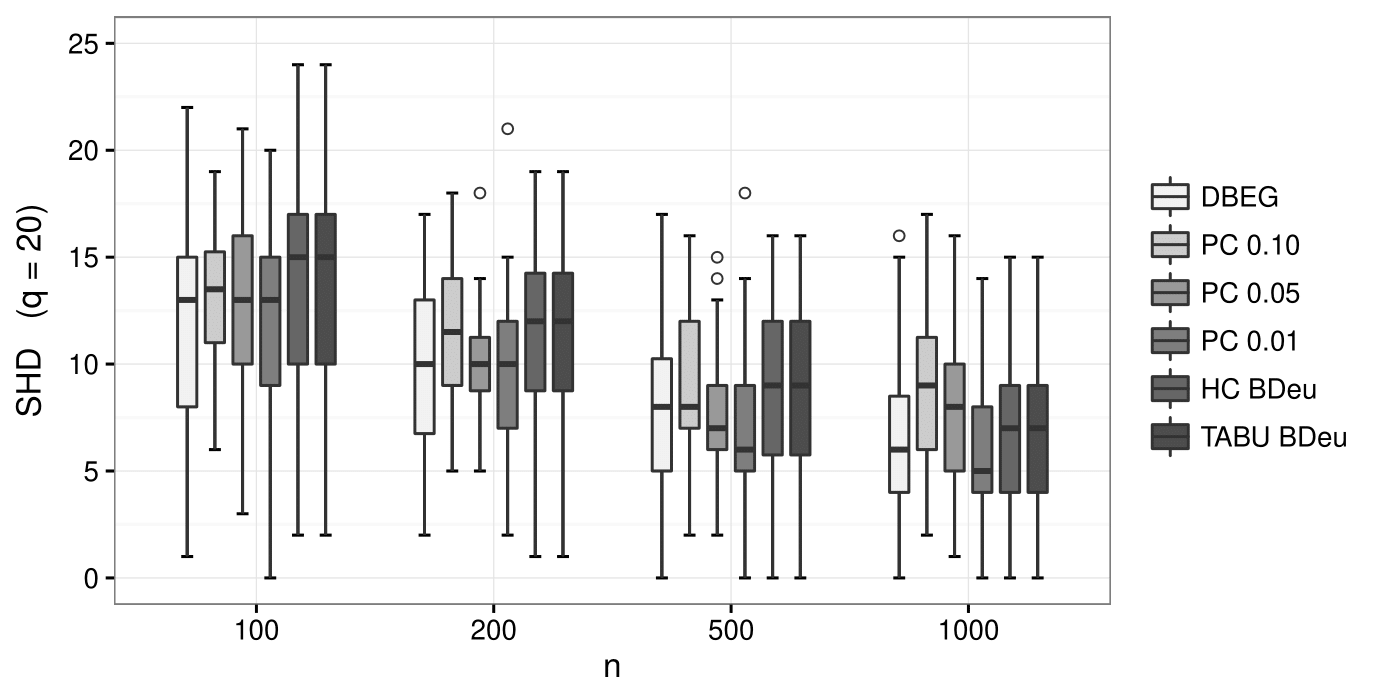}
			\includegraphics[scale=0.2]{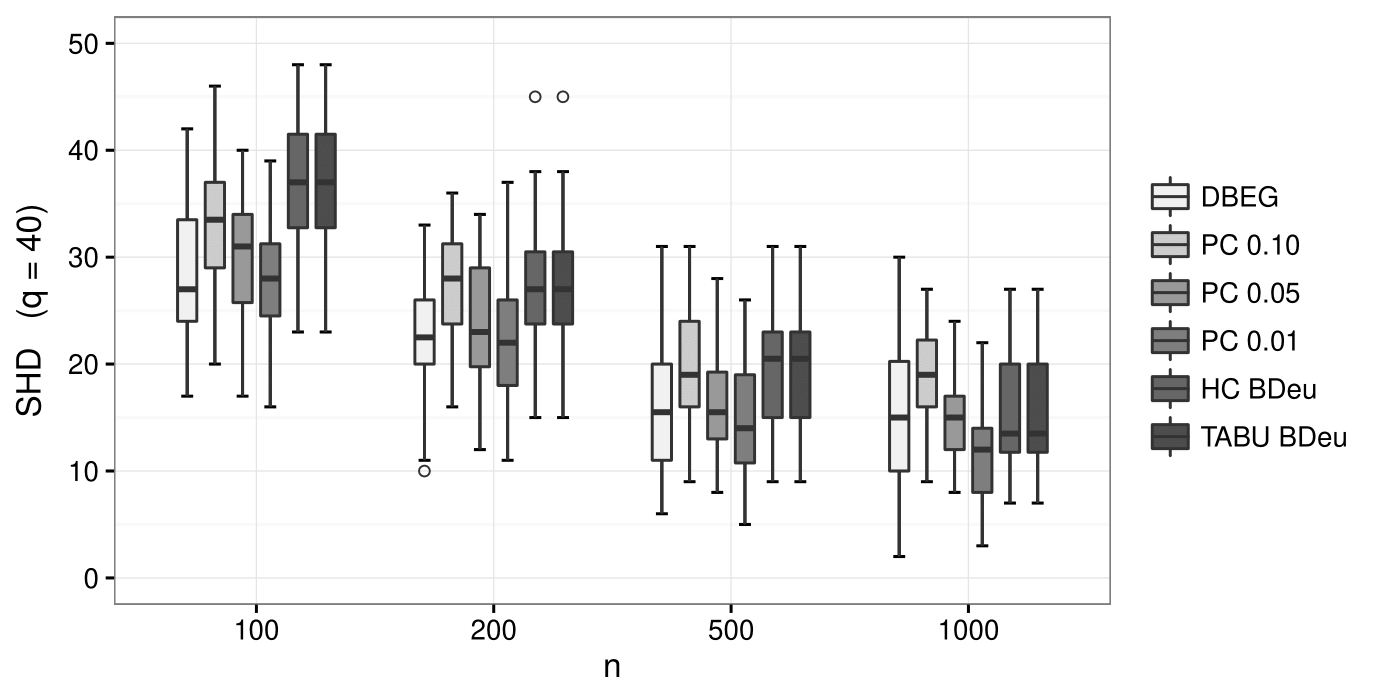} \\
			\caption{\small Simulations. Structural Hamming Distance (SHD) between true and estimated EG for number of nodes $q\in\{5,10,20,40\}$ and increasing samples sizes $n\in\{100,200,500,1000\}$. Methods under comparison are: our DBEG method, 
				the PC algorithm of \citet{Spir:Glym:Sche:2000}, implemented for significance levels $\alpha\in\{0.10,0.05,0.01\}$ (respectively PC 0.10, PC 0.05, PC 0.01), HC BDeu, a hill climbing greedy search with the BDeu score of \citet{Heckerman:1995}, and its modified version TABU BDeu (\citet{russell2002artificial}).}
			\label{fig:shds}
		\end{center}
	\end{figure*}

We evaluate the ability of each method in recovering the true EG structure in terms of Structural Hamming Distance (SHD) between true and estimated EG. The SHD represents the number of edge insertions, deletions or flips needed to transform
the estimated EG into the true one. Accordingly, lower values of SHD correspond to better
performances. Results are summarized in the box-plots of Figure \ref{fig:shds}, where each plot reports the distribution of SHD across the simulated datasets for a given value of $q\in\{5,10,20,40\}$ and increasing sample sizes $n\in\{100,200,500,1000\}$.
With regard to our method we consider as EG point estimate 
the projected median probability graph model (DBEG); see also Section \ref{sec:MCMC:posterior_model:inference}. All methods improve their performance as the sample size increases. Moreover, our DBEG method outperforms PC 0.10, PC 0.05, HC BDeu and TABU BDeu most of the times and remains
highly competitive with PC 0.01 under all scenarios.

For each scenario and method we also evaluate the performance in learning the structure of the true EG in terms of misspecification rate, specificity, sensitivity, precision
and Matthews correlation coefficient:
\be
&\mathrm{MISR} = \frac{FN+FP}{q(q-1)},
\quad \mathrm{SPE} = \frac{TN}{TN+FP},\\
&\mathrm{SEN} = \frac{TP}{TP+FN},
\quad \mathrm{PRE}  = \frac{TP}{TP+FP},\\
&\mathrm{MCC} = \frac{TP\cdot TN - FP \cdot FN}{\sqrt{(TP+FP)(TP+FN)(TN+FP)(TN+FN)}},
\ee
where $TP$, $TN$, $FP$, $FN$ are the numbers of true positives, true negatives, false positives and false negatives respectively. The four measures can be computed by comparing the true and estimated EG through the corresponding adjacency matrices, where an undirected edge $u-v$ is treated as the union of the two directed edges $u \rightarrow v$ and $u \leftarrow v$. With the exception of MISR, better performances correspond to higher values.

Results for number of nodes $q\in\{10,20\}$ are summarized in Tables \ref{tab:measures:q10} and \ref{tab:measures:q20}, where we compare our DBEG with the three versions of the PC algorithm, and with the score-based methods of HC BDeu and TABU BDeu.
In terms of specificity index (SPE), all methods are comparable. The superiority of DBEG and PC 0.01, relative to the other methods, stems from a higher precision (PRE) and higher Matthews correlation coefficient (MCC), the latter being more evident in the setting $q=20$ or for larger sample sizes. Moreover, there is no clear ranking between the three versions of the PC algorithm in terms of sensitivity index (SEN), while HC BDeu and TOTEM BDeum are indistinguishable.

\begin{table*}
	\centering
	\begin{tabular}{lcccccc}
		\hline
		\hline
		& & MISR & SPE & SEN & PRE & MCC \\ 
		\hline
		\hline
		\multirow{6}{1.8cm}{$n = 100$}
		& DBEG & 8.31 & 97.87 & 45.76 & 75.69 & 57.54 \\ 
		& PC 0.10 & 8.08 & 97.38 & 52.35 & 72.61 & 59.91 \\ 
		& PC 0.05 & 7.67 & 98.07 & 49.99 & 77.08 & 60.26 \\ 
		& PC 0.01 & 8.19 & 98.63 & 42.54 & 79.53 & 56.38 \\ 
		& HC BDeu & 8.92 & 98.32 & 37.00 & 74.91 & 51.25 \\ 
		& TABU BDeu & 8.92 & 98.32 & 37.00 & 74.91 & 51.25 \\ 
		\hline
		\multirow{6}{1.8cm}{$n = 200$}
		& DBEG & 6.14 & 98.75 & 61.21 & 82.97 & 69.33 \\ 
		& PC 0.10 & 7.17 & 97.11 & 62.39 & 73.95 & 65.89 \\ 
		& PC 0.05 & 6.31 & 97.94 & 63.80 & 81.29 & 69.99 \\ 
		& PC 0.01 & 6.31 & 98.60 & 58.71 & 85.64 & 68.69 \\ 
		& HC BDeu & 7.58 & 99.08 & 42.85 & 88.24 & 59.25 \\ 
		& TABU BDeu & 7.58 & 99.08 & 42.85 & 88.24 & 59.25 \\ 
		\hline
		\multirow{6}{1.8cm}{$n = 500$} 
		& DBEG & 4.69 & 98.44 & 73.17 & 86.55 & 77.90 \\ 
		& PC 0.10 & 5.78 & 97.3 & 72.79 & 78.61 & 73.55 \\ 
		& PC 0.05 & 5.33 & 97.94 & 71.77 & 83.35 & 75.33 \\ 
		& PC 0.01 & 4.69 & 98.82 & 70.63 & 89.9 & 77.75 \\ 
		& HC BDeu & 7.33 & 98.6 & 48.41 & 83.43 & 61.8 \\ 
		& TABU BDeu & 7.33 & 98.6 & 48.41 & 83.43 & 61.8 \\ 
		\hline
		\multirow{6}{1.8cm}{$n = 1000$}
		& DBEG & 4.22 & 98.16 & 79.66 & 85.94 & 83.20 \\ 
		& PC 0.10 & 5.33 & 97.36 & 75.37 & 78.41 & 74.97 \\ 
		& PC 0.05 & 4.31 & 98.19 & 78.29 & 85.51 & 80.15 \\ 
		& PC 0.01 & 3.72 & 98.86 & 77.84 & 90.36 & 82.23 \\ 
		& HC BDeu & 7.03 & 98.45 & 51.84 & 83.12 & 63.84 \\ 
		& TABU BDeu & 7.03 & 98.45 & 51.84 & 83.12 & 63.84 \\ 
		\hline
		\hline
	\end{tabular}
	\vspace{0.2cm}
	\caption{\label{tab:measures:q10} Simulations. Misspecification rate (MISR), specificity (SPE), sensitivity (SEN), precision (PRE) and Matthews correlation coefficient (MCC) averaged over 40 simulations for number of nodes $q = 10$ and sample size $n\in\{100,200,500,1000\}$ for each method under comparison.}
\end{table*}

\begin{table*}
	\centering
	\begin{tabular}{lcccccc}
		\hline
		\hline
		& & MISR & SPE & SEN & PRE & MCC \\ 
		\hline
		\hline
		\multirow{6}{1.8cm}{$n = 100$}
		& DBEG & 4.17 & 99.05 & 37.75 & 70.17 & 50.59 \\ 
		& PC 0.10 & 4.48 & 98.47 & 42.45 & 59.71 & 49.89 \\ 
		& PC 0.05 & 4.49 & 98.62 & 40.06 & 61.07 & 49.10 \\ 
		& PC 0.01 & 4.11 & 99.13 & 38.39 & 70.95 & 51.46 \\ 
		& HC BDeu & 4.67 & 98.58 & 36.12 & 58.31 & 45.74 \\ 
		& TABU BDeu & 4.67 & 98.58 & 36.12 & 58.31 & 45.74 \\ 
		\hline
		\multirow{6}{1.8cm}{$n = 200$}
		& DBEG & 3.34 & 98.92 & 55.87 & 75.64 & 64.03 \\ 
		& PC 0.10 & 3.78 & 98.49 & 55.20 & 66.22 & 59.87 \\ 
		& PC 0.05 & 3.47 & 98.85 & 54.82 & 71.90 & 62.05 \\ 
		& PC 0.01 & 3.37 & 99.22 & 50.74 & 77.64 & 61.97 \\ 
		& HC BDeu & 4.09 & 98.88 & 42.21 & 68.10 & 53.10 \\ 
		& TABU BDeu & 4.09 & 98.88 & 42.21 & 68.10 & 53.10 \\ 
		\hline
		\multirow{6}{1.8cm}{$n = 500$} 
		& DBEG & 2.67 & 99.02 & 67.42 & 80.51 & 72.82 \\ 
		& PC 0.10 & 2.81 & 98.64 & 71.91 & 74.24 & 72.22 \\ 
		& PC 0.05 & 2.49 & 98.94 & 72.31 & 78.96 & 74.69 \\ 
		& PC 0.01 & 2.30 & 99.33 & 69.51 & 85.19 & 76.04 \\ 
		& HC BDeu & 3.6 & 98.98 & 49.63 & 74.38 & 59.92 \\ 
		& TABU BDeu & 3.6 & 98.98 & 49.63 & 74.38 & 59.92 \\ 
		\hline
		\multirow{6}{1.8cm}{$n = 1000$}
		& DBEG & 2.32 & 98.96 & 78.06 & 81.06 & 80.48 \\ 
		& PC 0.10 & 2.72 & 98.46 & 76.54 & 73.59 & 74.17 \\ 
		& PC 0.05 & 2.36 & 98.83 & 77.01 & 78.62 & 77.02 \\ 
		& PC 0.01 & 1.93 & 99.35 & 76.15 & 86.55 & 80.26 \\ 
		& HC BDeu & 3.09 & 99.17 & 55.63 & 79.89 & 65.70 \\ 
		& TABU BDeu & 3.09 & 99.17 & 55.63 & 79.89 & 65.70 \\ 
		\hline
		\hline
	\end{tabular}
	\vspace{0.2cm}
	\caption{\label{tab:measures:q20} Simulations. Misspecification rate (MISR), specificity (SPE), sensitivity (SEN), precision (PRE) and Matthews correlation coefficient (MCC) averaged over 40 simulations for number of nodes $q = 20$ and sample size $n\in\{100,200,500,1000\}$ for each method under comparison.}
\end{table*}


Simulated datasets were obtained by generating first (latent) continuous multivariate Gaussian observations as in \eqref{eq:generating:process} that were subsequently discretized to obtain binary data by fixing a threshold $\gamma_j = 0$; see Equation \eqref{eq:discretize:gaussian}.
The zero threshold, coupled with the assumption $\mu_j=0$ in \eqref{eq:generating:process} which implies a marginal mean equal to zero for each latent $Z_j$, results in a collection of categorical variables whose levels are well balanced, meaning that $P(Y_j=1) = P(Y_j=0) = 0.5$ for each $j=1,\dots,q$.
In the following we relax this assumption by drawing each $\gamma_j$ uniformly in the interval $[0,1]$. As a consequence, each so-obtained dataset exhibits an excess of zeros, since now $P(Y_j=1)\le 0.5$ with a lower bound which depends on the marginal variance of each latent $Z_j$ (in our simulation settings approaching $0.15$ in the ``worst" case where $\gamma_j=1$).

\begin{figure*}
	\begin{center}
		\includegraphics[scale=0.2]{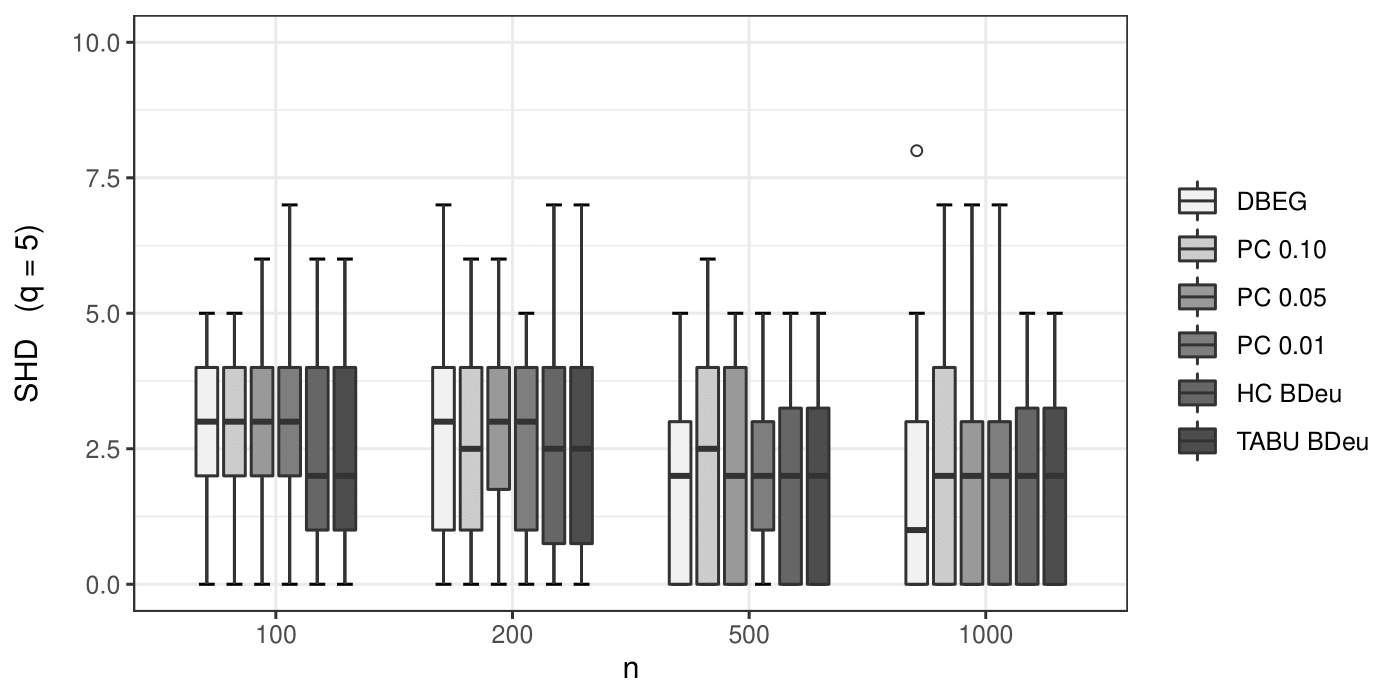}
		\includegraphics[scale=0.2]{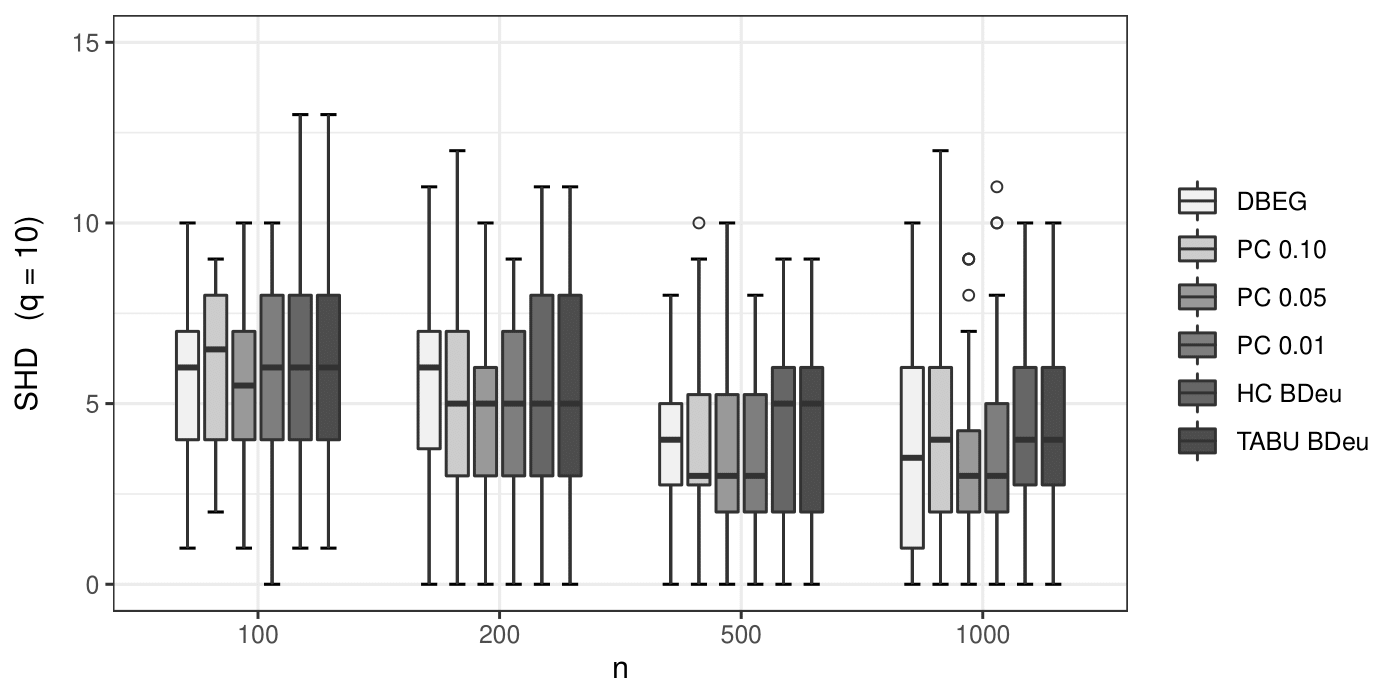} \\
		\includegraphics[scale=0.2]{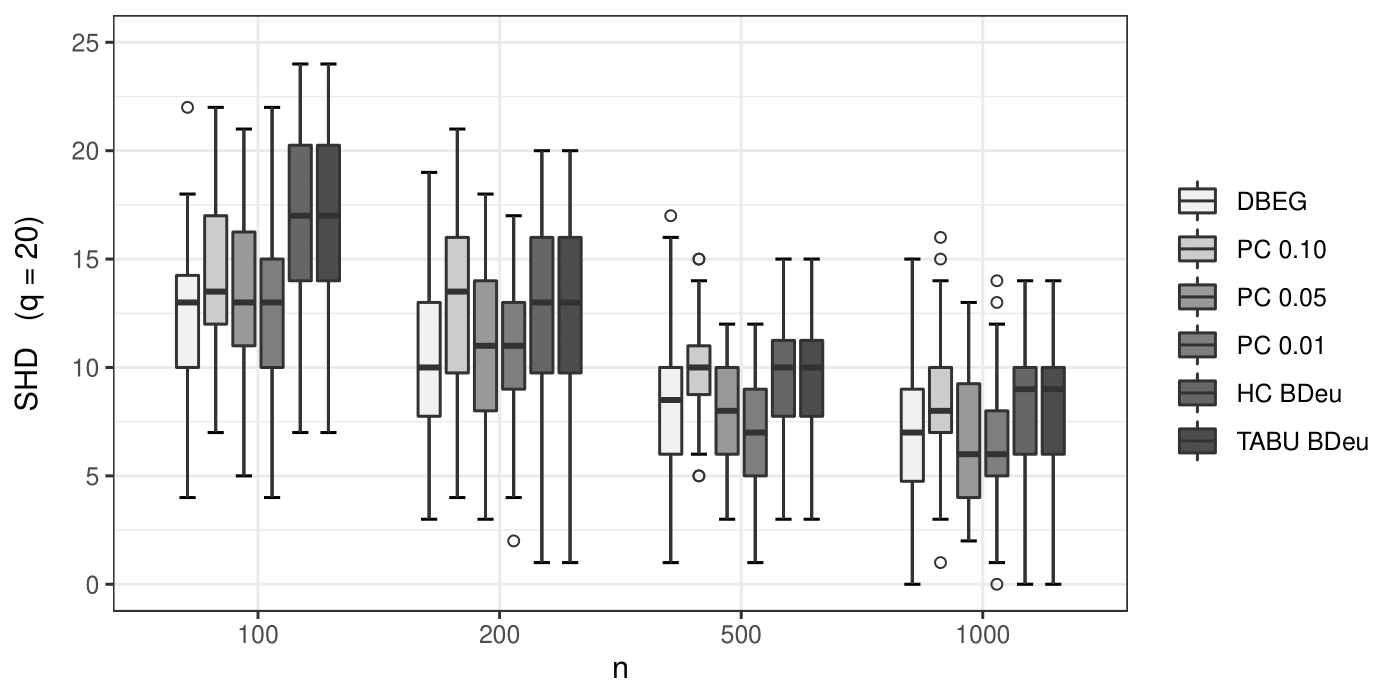}
		\includegraphics[scale=0.2]{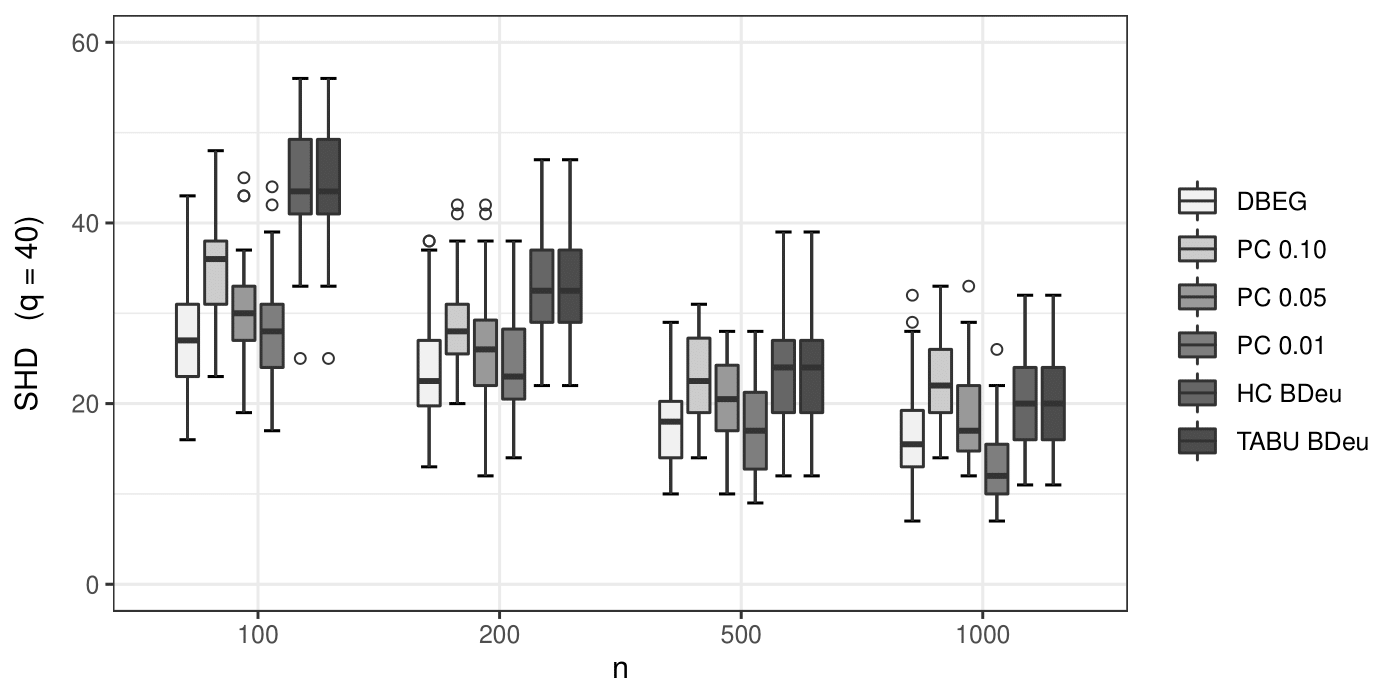} \\
		\caption{\small Simulations (unbalanced setting). Structural Hamming Distance (SHD) between true and estimated EG for number of nodes $q\in\{5,10,20,40\}$ and increasing samples sizes $n\in\{100,200,500,1000\}$. Methods under comparison are: our DBEG method, 
			the PC algorithm of \citet{Spir:Glym:Sche:2000}, implemented for significance levels $\alpha\in\{0.10,0.05,0.01\}$ (respectively PC 0.10, PC 0.05, PC 0.01), HC BDeu, a hill climbing greedy search with the BDeu score of \citet{Heckerman:1995}, and its modified version TABU BDeu (\citet{russell2002artificial}).}
		\label{fig:shds:unbalanced}
	\end{center}
\end{figure*}

Simulation results are reported in Figure \ref{fig:shds:unbalanced} where the box-plots summarize the distribution of SHD for values of $q\in\{5,10,20,40\}$ and $n\in\{100,200,500,10000\}$ for each method under comparison.
Results are very similar to those obtained under the ``balanced" setting where $\gamma_j=0$, with our DBEG approach outperforming the two BDeu-based methods and being competitive with PC in most of the settings, in particular for scenarios characterized by moderate sample sizes.
The same behaviour was observed for each of the five indexes in Tables \ref{tab:measures:q10}-\ref{tab:measures:q20} that we do not include for brevity.

\section{Real data analyses}\label{sec:app}
\subsection{Alarm data}
We apply our method to the ALARM dataset presented in \citet{Beinlich:et:al:1989}.
ALARM (A Logical Alarm Reduction Mechanism) is an alarm message system for patient monitoring based on a diagnostic tool.
From a graphical model viewpoint, ALARM consists of a \textit{belief network}, a DAG describing dependence relationships between three types of categorical variables: 8 diagnoses (at the top level of the network), 13 intermediate variables and 16 findings (clinical outcomes). A number of observations $n=20000$ are measured on each of the $q=37$ categorical variables.
Of these, $13$ variables are binary, while the others have a number of levels equal to $3$ or $4$ ($17$ and $7$ variables respectively).
The objective of the original study was to estimate parameters (i.e. conditional probabilities) of interest as a diagnostic tool for patient monitoring, given a known DAG structure with
46 directed edges; see also \citet[Figure 1]{Beinlich:et:al:1989}.

\begin{figure*}
	\begin{center}
		\includegraphics[scale=0.2]{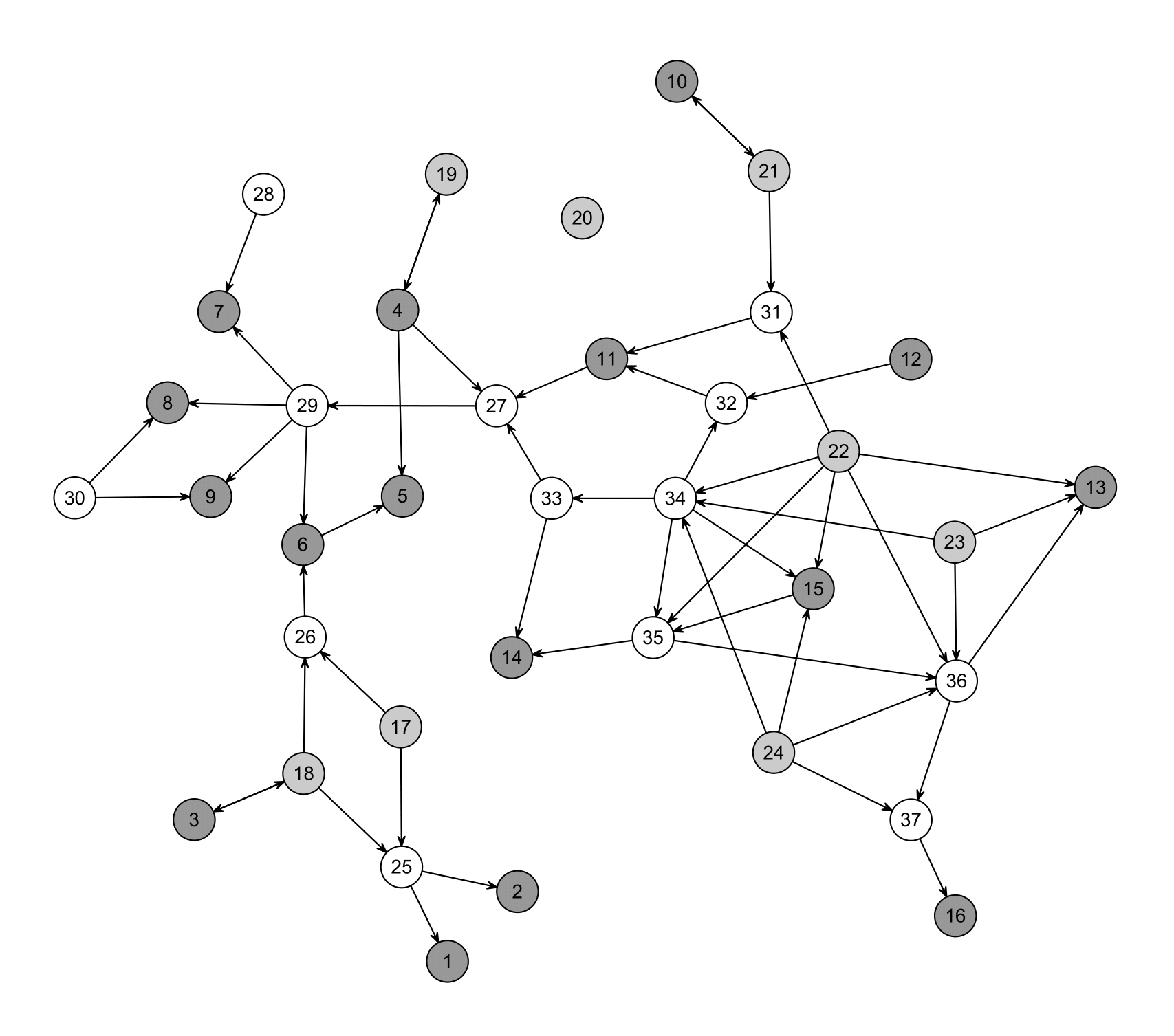}
		\caption{\small ALARM data. Estimated EG (maximum a posteriori and median probability graph model) obtained under DBEG.}
		\label{fig:alarm:graph}
	\end{center}
\end{figure*}

\begin{figure*}
	\begin{center}
		\includegraphics[scale=0.1]{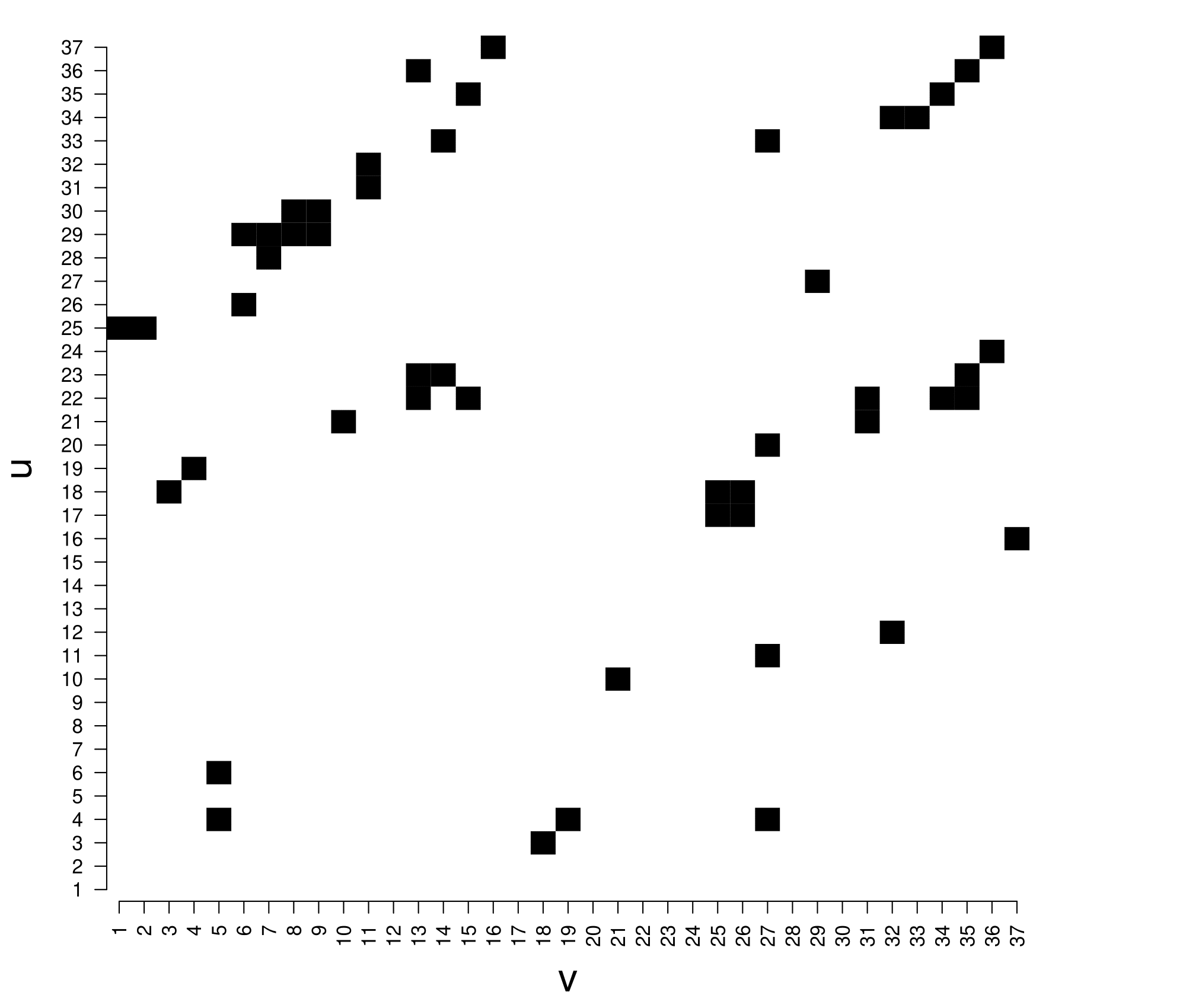}
		\includegraphics[scale=0.1]{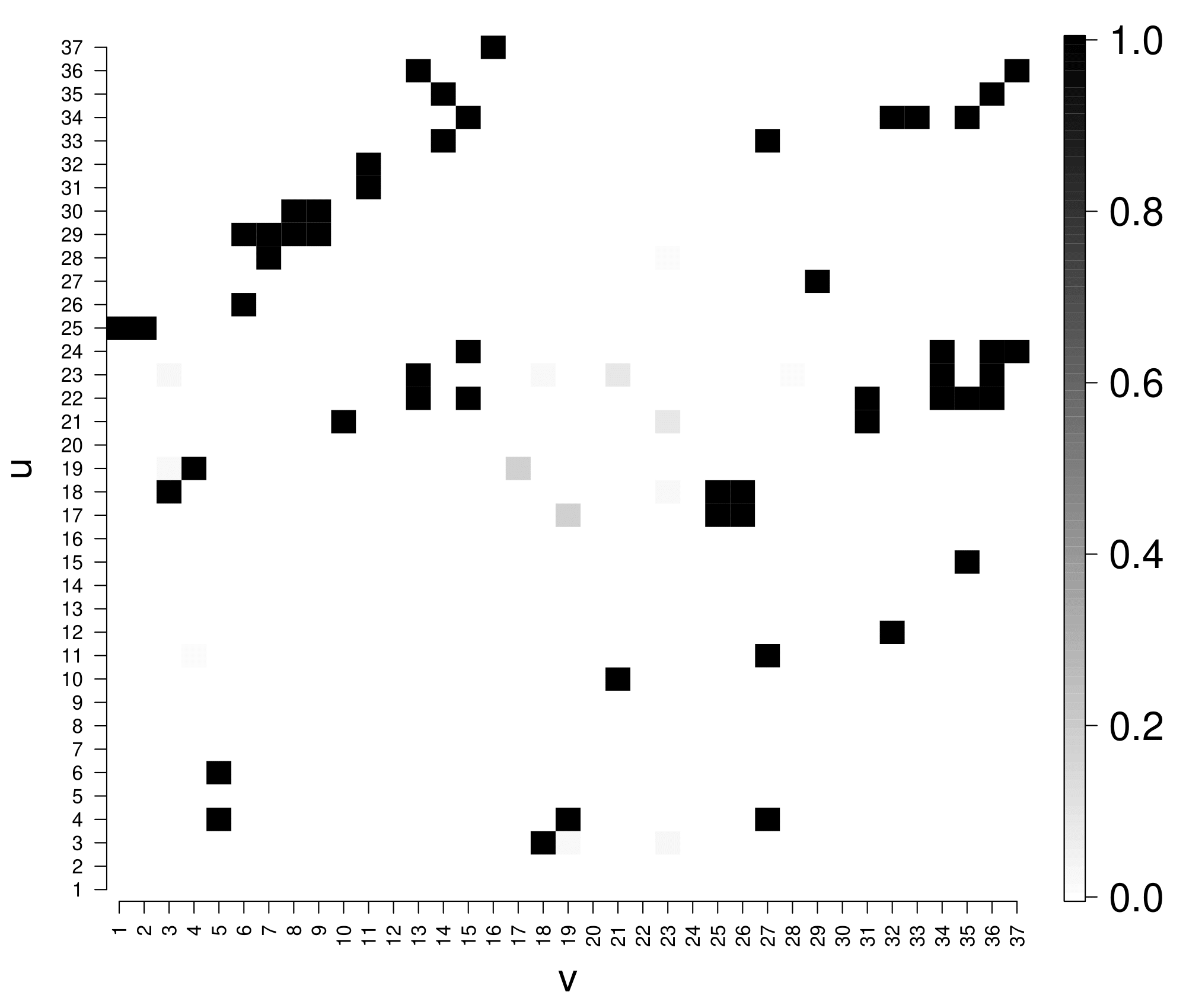}
		\caption{\small ALARM data. Comparison between the original DAG assumed in \citet{Beinlich:et:al:1989} (left), with black dots in correspondence of edges, and estimated posterior probabilities of edge inclusion from our DBEG method (right).}
		\label{fig:alarm:heat}
	\end{center}
\end{figure*}

On the other hand, we account for
uncertainty in the data-generating graphical model and we implement our methodology to learn an EG structure. This can be compared with the clinically justified DAG graphical structure assumed as known in the original study.
We run $T = 50000$ iterations of DBEG, by fixing a prior probability of edge inclusion $\pi=0.02$ to favour sparsity and hyperparameter $a(y_{\tau}\g r)=1/l_{\tau}$ in the Dirichlet prior \eqref{eq:prior:dir:complete}.
The MCMC output estimates a posterior distribution on the EG space which is highly concentrated, with a single EG model assigned a posterior probability of about $70\%$. Therefore, the maximum a posteriori and the (projected) median probability graph models coincide (Figure \ref{fig:alarm:graph}).
Estimated posterior probabilities of edge inclusion, as in Equation \eqref{eq:posterior:edge:inclusion}, are summarized in the (left-side) heat map of Figure \ref{fig:alarm:heat} and confirm the low variability of the EG posterior distribution. All edges included in the EG estimate of Figure \ref{fig:alarm:graph} have indeed a posterior probability close to one.
Few exceptions are represented by edges $17-19$, $21-23$ and $3-23$, whose posterior probabilities however does not exceed the threshold for edge inclusion. In the same figure we provide a comparison with the EG implied by the DAG model assumed in \citet{Beinlich:et:al:1989}, here represented as a heat map with black dots in correspondence of edges.
The two plots reveal strong similarities between the EG structures, since they differ by 16 edges over 46 and 51 edges respectively included in the two graphs.


\subsection{Voting records}
In this section we analyze the voting records from the 1984 United Stated Congress.
The dataset includes votes for each of the $n=434$ U.S. House of Representatives Congressmen on sixteen key votes, identified by the Congressional Quarterly Almanac, on religion, immigration, crime, education, and other relevant subjects. Each of the $q=16$ (categorical) answers takes value in $\{yes,no,NA\}$, with $NA$ in case of missing response. The data are publicly available at \url{https://archive.ics.uci.edu/}.
In the following we also distinguish between democratic and republican Congressmen by considering two datasets with $n_D=267$ and $n_R=167$ observations respectively.
Our method is then applied independently to each dataset, by fixing the number of MCMC iterations $T=30000$, the prior probability of edge inclusion $\pi=5\%$ and the hyperparameter $a(y_{\tau}\g r)=1/l_{\tau}$ in the Dirichlet prior \eqref{eq:prior:dir:complete}.

\begin{figure*}
	\begin{center}
		\includegraphics[scale=0.17]{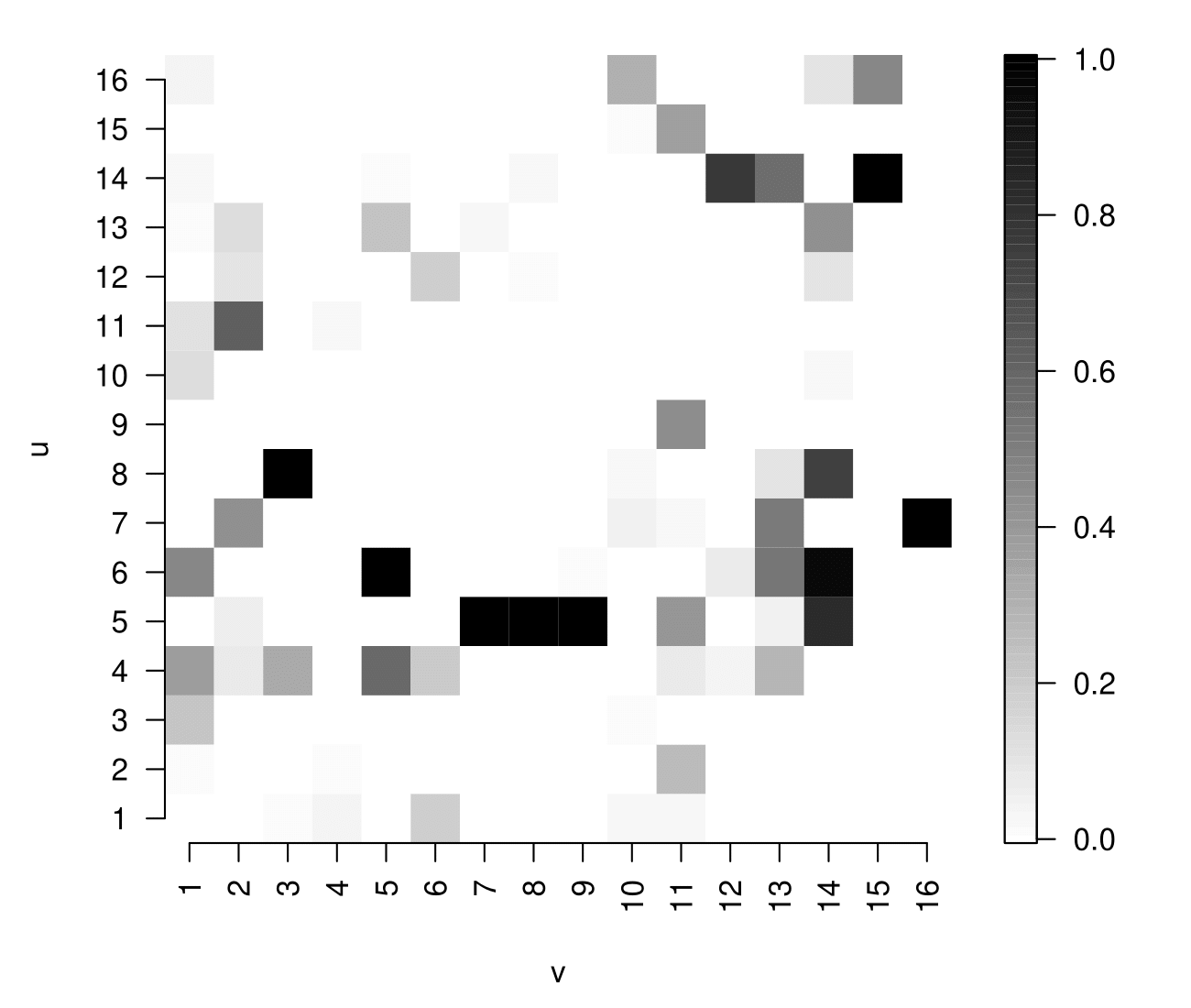}
		\includegraphics[scale=0.17]{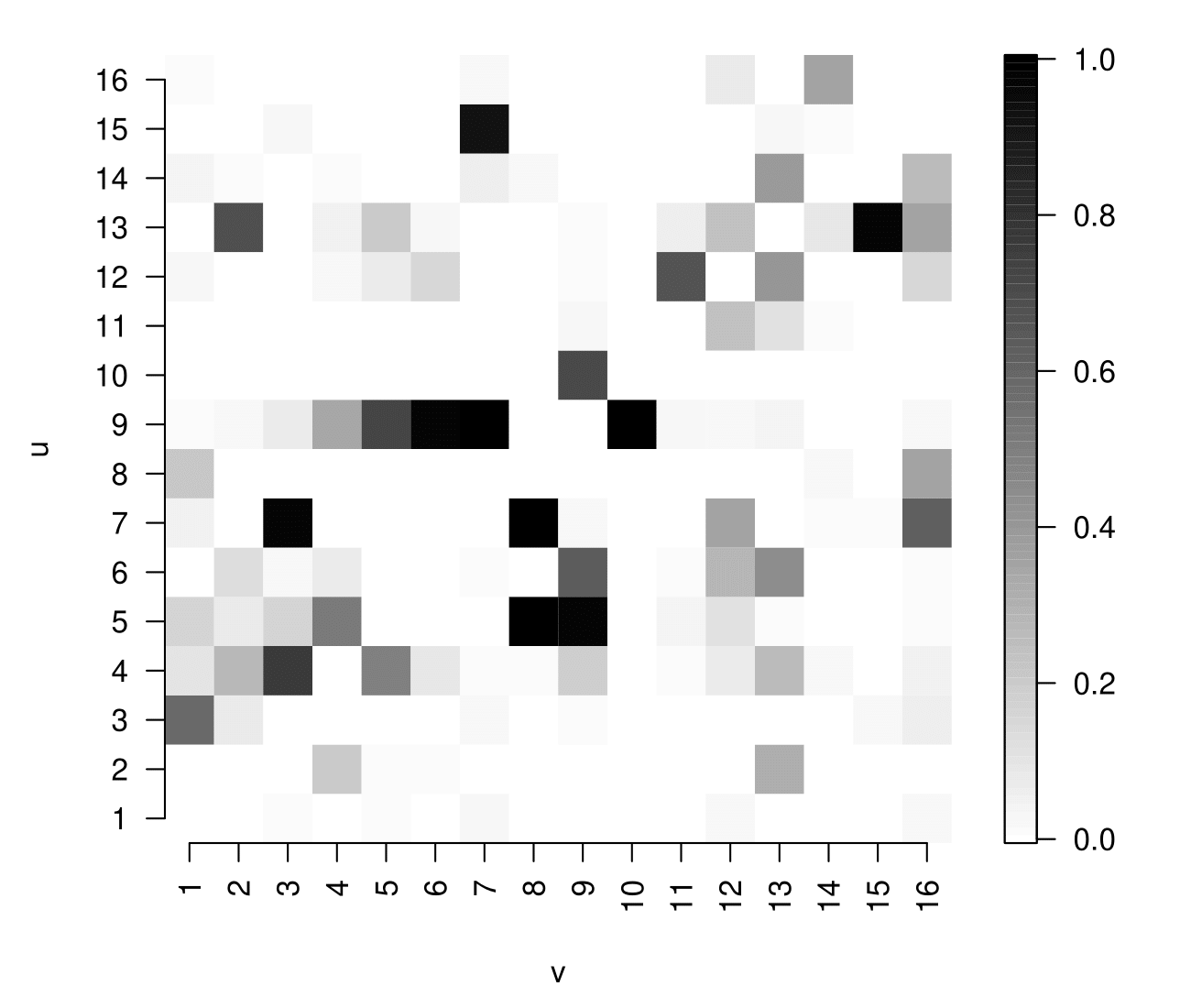}
		\caption{\small Voting records. Heat maps with estimated posterior probabilities of edge inclusion obtained from our DBEG method for the two groups: democratic (left) and republican (right).}
		\label{fig:votes:heat}
	\end{center}
\end{figure*}

\begin{figure*}
	\begin{center}
		\includegraphics[scale=0.18]{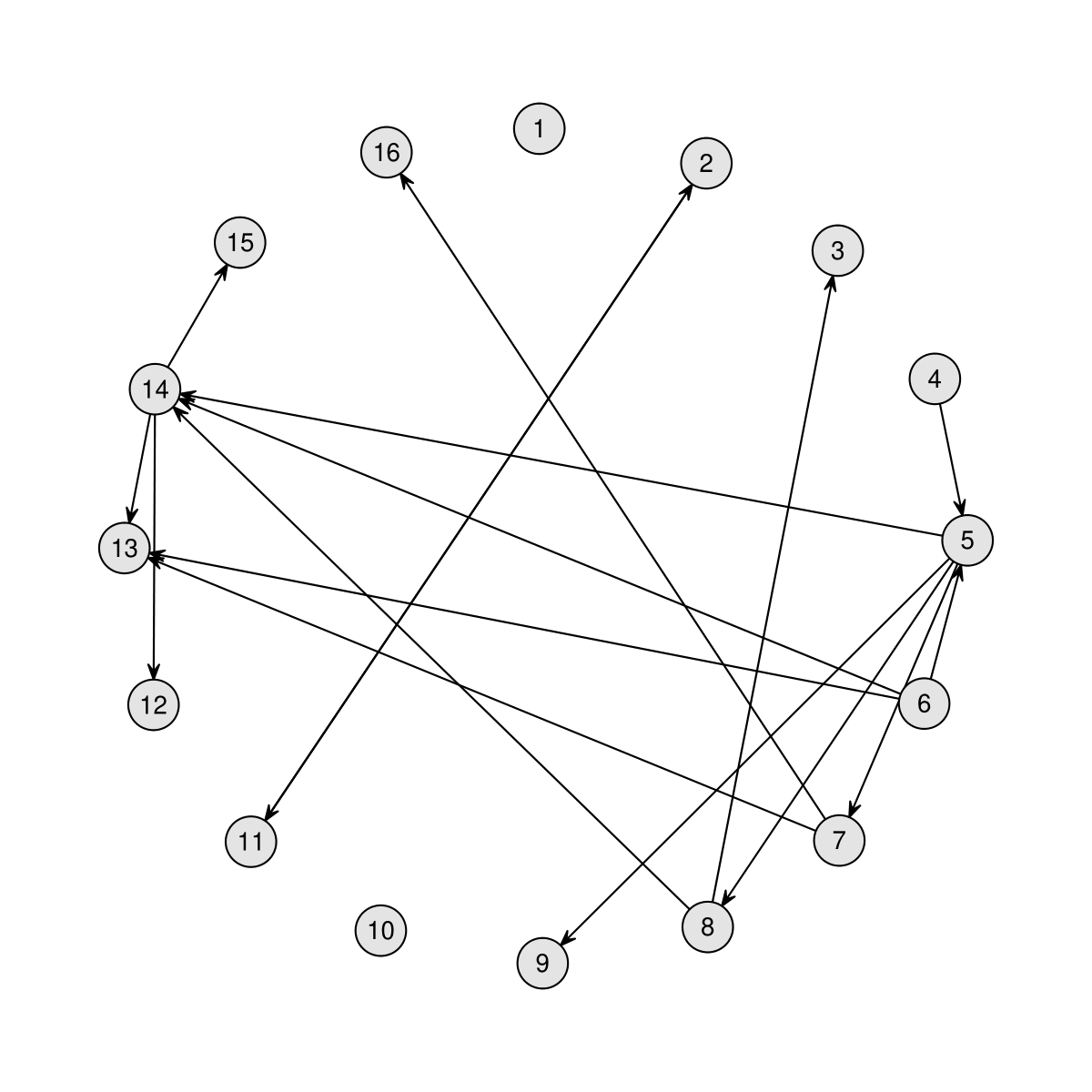}
		\includegraphics[scale=0.18]{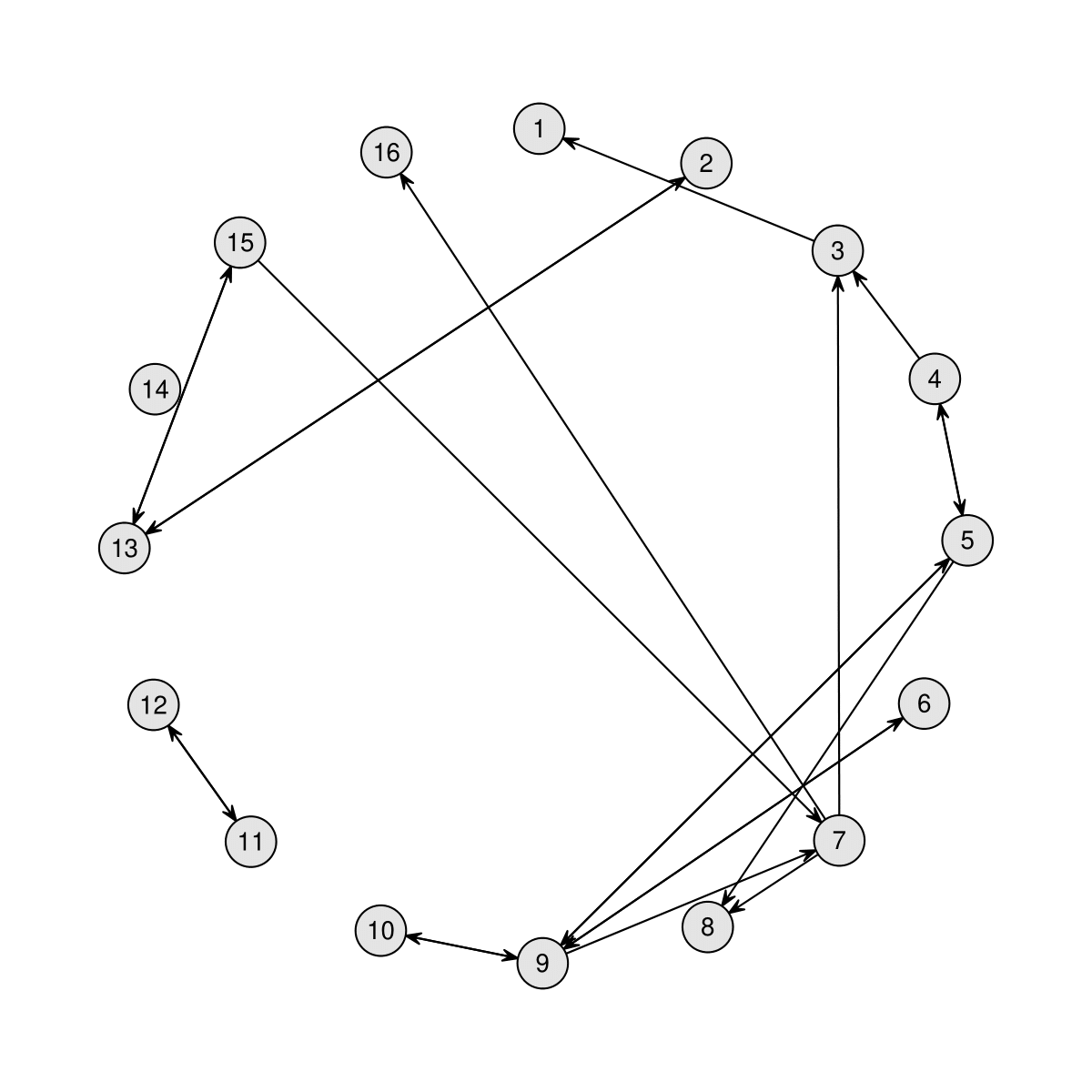}
		\caption{\small Voting records. Estimated EG (projected median probability graph model) obtained with DBEG for the two groups: democratic (left) and republican (right).}
		\label{fig:votes:graph}
	\end{center}
\end{figure*}

Differently from the previous application, the posterior distribution over the EG space exhibits larger variability, possibly related to the more moderate group sample sizes.
This is also apparent from Figure \ref{fig:votes:heat} which summarizes
the estimated posterior probabilities of edge inclusion under each group.
In addition, the two plots (democratic and republican) reveal strong differences, as evident from the estimated graphs in Figure \ref{fig:votes:graph}.
Few exceptions of similarity are the (directed) links between
5 (el-salvador-aid) and
8 (aid-to-nicaraguan-contras),
7 (anti-satellite-test-ban) and
16 (export-administration-act-south-africa), common to the two groups.

\section{Conclusions and further directions}
\label{sec:discussion}

We propose a Bayesian method for learning the conditional dependence structures of multivariate categorical data that we represent through a Directed Acyclic Graph (DAG). To account for different DAGs encoding the same set of dependencies (Markov equivalent DAGs), and to avoid hyperprior specifications that lead to undesirable properties of the marginal likelihood, our methodology directly learns the essential graph (EG) representative of a DAG equivalence class. Following the method of \citet{Geig:Heck:2002} for parameter prior construction, we derive a closed-form expression of the EG marginal likelihood, in accordance with the graph-driven likelihood decomposition, and study related asymptotic properties. These developments serve a proposed MCMC sampler on the EG space, that we apply to simulated data in comparison with benchmarks and on two real datasets.

With interventional data subject to exogenous perturbations or randomized experiments, the marginal likelihood can still be factorized according to the conditional independence structure implied by the graph (\citet{Pear:2000}, \citet{hauser2015jointly}). Interventional Markov equivalence classes (\citet{He:etal:2008}, \citet{Haus:Buhl:2012}) preserve the characterization as chain graphs with decomposable chain components, but they  
constitute a finer partition of the DAG space, relative to their observational counterpart,
and therefore improve the identifiability of the true data generating DAG. A generalization of the proposed setting to interventional categorical data is of interest, and would be based, following \citet{Caste:Cons:2019}, on the extension of the EG marginal likelihood to I-EGs (interventional essential graphs) and of the Markov chain of \citet{He:etal:2013} to the I-EG space.

Also, the US voting datasets of democratics and republicans were analyzed separately, assuming distinct graphical structures (one for each group) that accordingly were estimated independently. Alternatively, one could analyze them jointly to exploit potential shared features among groups. Joint structural learning for multiple Gaussian undirected graphs is carried out in \citet{Pete:etal:2015}, through a Markov random field prior that encourages common edges, and a spike-and-slab prior on network relatedness parameters. Their framework has been extended to Gaussian EGs in \citet{Castelletti2020} and, along the same dimension, an extension of our methodology to infer multiple categorical EGs is feasible and under investigation.

\section*{Appendix: Graph notation}
\label{sec:appendix}
A graph $\G$ is a pair $(V,E)$ where $V=\{1,\dots,q\}$ is a set of vertices (or nodes) and $E \subseteq V \times V$ a set of edges (or arcs).
Nodes are associated to variables, while edges are used to represent direct interactions between variables.
Let $u,v \in V$, $u \ne v$ be two nodes. We say that $\G$ contains the directed edge $u \rightarrow v$ if and only if $(u,v) \in E$ and $(v,u) \notin E$. If instead both $(u,v) \in E$ and $(v,u) \in E$, then $\G$ contains the undirected edge $u - v$.
Accordingly, we say that $\G$ is an undirected (directed) graph if it contains only undirected (directed) edges; in addition, $\G$ is \textit{partially directed} if it contains at least one directed edge.

Two vertices $u,v$ are adjacent if they are connected by an edge (directed or undirected).
In addition, we call $u$ a \textit{neighbor} of $v$ if $u - v$ is in $\G$ and denote the neighbor set of $v$ as $\neigh_{\G}(v)$; the common neighbor set of $u$ and $v$ is then $\neigh_{\G}(u,v)=\neigh_{\G}(u) \cap \neigh_{\G}(v)$.
We say that $u$ is a \textit{parent} of $v$ and that $v$ is a \textit{child} of $u$ if $u\rightarrow v$ is in $\G$. The set of all parents of $u$ in $\G$ is then denoted by $\pa_{\G}(u)$.
A sequence of nodes $\{v_0,v_1,\dots,v_k\}$ where $v_0=v_k$ and $v_{j-1} - v_j$ or $v_{j-1} \rightarrow v_j$ for all $j=1,\dots,k$ is called a \textit{cycle}.
A cycle is directed (undirected) if it contains only directed (undirected) edges; conversely we call it a partially-directed cycle. A graph with only directed edges is called a directed acyclic graph (DAG) if it does not contain cycles.
For any subset $A\subseteq V$ we denote with $\G_{A}=(A,E_{A})$ the \textit{subgraph} of $\G$ induced by $A$, where $E_{A}=E \cap (A\times A)$.
A (sub)graph is complete if its vertices are all adjacent.

\begin{figure}
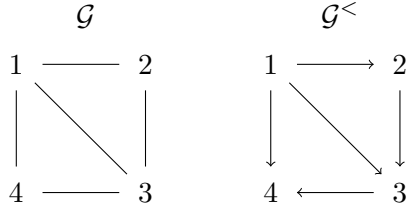

	\begin{center}
		\begin{tabular}{cccc}
			$\G$ & & $\G^{<}$ \\
			{\normalsize
				\tikz \graph [no placement, math nodes, nodes={circle}]
				{
					1[x=0,y=0] -- {2[x=1.7,y=-0],4[x=0,y=-1.7],3[x=1.7,y=-1.7]},
					2[x=1.7,y=0] -- 3[x=1.7,y=-1.7],
					3[x=1.7,y=-1.7] -- 4[x=0,y=-1.7]};
			}
			& &
			{\normalsize
				\tikz \graph [no placement, math nodes, nodes={circle}]
				{
					1[x=0,y=0] -> {2[x=1.7,y=-0],4[x=0,y=-1.7],3[x=1.7,y=-1.7]},
					2[x=1.7,y=0] -> 3[x=1.7,y=-1.7],
					3[x=1.7,y=-1.7] -> 4[x=0,y=-1.7]};
			}
		\end{tabular}
	\end{center}
	\caption{\small A decomposable graph $\G$ on the set of vertices $V=\{1,2,3,4\}$; the cycle $\{1,2,4,3\}$ of length $l=4$ contains the chord $1 - 3$. $\G$ has the perfect sequence of cliques $\{C_1,C_2\}$, with $C_1=\{1,2,3\},C_2=\{1,3,4\}$ and set of separators $\mathcal{S}=\{S_2\}$, $S_2=\{1,3\}$. $\G^{<}$ is the perfect directed version of $\G$.}
	\label{fig:dec:perfect}
\end{figure}

We now focus on a particular class of undirected graphs, namely \textit{decomposable} graphs (also called \textit{chordal} or \textit{triangulated}).
Specifically, we say that an undirected (sub)graph is decomposable if every cycle of length $l\ge 4$ has a \textit{chord}, that is two nonconsecutive adjacent vertices. For a decomposable graph $\G$, a complete subset that is maximal with respect to inclusion is called a \textit{clique}. Let $\mathcal{C} = \{C_1,\dots, C_K\}$ be a \textit{perfect} sequence of cliques. Let also $H_k = C_1 \cup \dots \cup C_k$, for $k=2,\dots,K$. We can then construct the set of \textit{separators} $\mathcal{S} = \{S_2,\dots,S_K\}$ where $S_k = C_k \cap H_{k-1}$; see also Figure \ref{fig:dec:perfect}.
It can be shown \citep[p.18]{Laur:1996} that each decomposable graph can be uniquely represented by its set of cliques and separators.
Most importantly, for each decomposable graph one can obtain a \textit{perfect numbering} of its vertices \citep{Laur:1996} and then a \textit{perfect directed version} $\G^{<}$ by directing its edges from lower to higher numbered vertices; see also Figure \ref{fig:dec:perfect}.

A partially directed graph with no partially-directed cycles is called a \textit{chain graph} (CG) or simply \textit{partially directed acyclic graph} (PDAG). For a chain graph $\G$ we call \textit{chain component} $\tau \subseteq V$ a set of nodes that are joined by an undirected path and denote the set of chain components of $\G$ by $\T$.
A subgraph of the form $u \rightarrow z \leftarrow v$, where there are no edges between $u$ and $v$, is called a \textit{v-structure} (or \textit{immorality}).
The \textit{skeleton} of a graph $\G$ is the undirected graph on the same set of vertices obtained by removing the orientation of all its edges.
Finally, a consistent extension of a PDAG $\G$ is a DAG on the same underlying set of edges, with the same orientations on the directed edges of $\G$ and the same set of \textit{v}-structures \citep{Dor:Tars:1992}.



%
%

\bibliographystyle{biometrika}       
\bibliography{biblio}   

\end{document}